%% file: manuscript.tex
\documentclass[authoryear,preprint]{elsarticle}
\usepackage{natbib}

\usepackage{graphicx}
\usepackage{wrapfig}
\usepackage{amssymb}
\usepackage{amsmath}
\usepackage{amsthm}
\usepackage{amsmath}
\usepackage{balance}
\usepackage{hyperref}
\usepackage{float}
\usepackage[export]{adjustbox}
\usepackage[normalem]{ulem}

\usepackage[symbol]{footmisc} 

\def\tsc#1{\csdef{#1}{\textsc{\lowercase{#1}}\xspace}}
\tsc{AU}
\newcommand{\sts}[1]{{#1}}

\begin{document}

\begin{frontmatter}

\title{Molecular dynamics study of the role of anisotropy in radiation-driven embrittlement$^\dagger$}

\author[IPPT]{H. Mousavi}

\author[IPPT]{S. Stupkiewicz}

\author[IPPT]{A. Ustrzycka\corref{cor1}}
\ead{austrzyc@ippt.pan.pl}
\cortext[cor1]{Corresponding author}

\address[IPPT]{Institute of Fundamental Technological Research, Polish Academy of Sciences,\\
Pawi\'nskiego 5B, 02-106 Warsaw, Poland.}

\begin{abstract}
This study investigates the influence of crystallographic orientation on fracture behavior and the resulting mechanical anisotropy in a $\rm Fe_{55}Ni_{19}Cr_{26}$ alloy crystal containing radiation-induced defects, using molecular dynamics (MD) simulations. Crack propagation is analyzed in irradiated samples with three selected high-symmetry crystallographic orientations to show how radiation-induced defects modify local deformation mechanisms and amplify mechanical anisotropy. The investigation focuses on the anisotropic nature of the ductile-to-brittle transition (DBT) driven by radiation-induced defects by simulating fracture behavior under tensile loading. Fracture resistance is quantitatively evaluated using a traction--separation (T--S) approach to extract the atomic-scale fracture energy under realistic defect conditions.  
The results reveal a strong crystallographic orientation dependence in the evolution of deformation and fracture behavior during DBT. The crystal lattice orientation governs dislocation activity and defect interactions, which in turn regulate local plasticity mechanisms, strain localization, slip system activation, and fracture resistance, thereby driving the development and enhancement of mechanical anisotropy in irradiated materials.  
It is further shown that radiation-induced embrittlement cannot be explained solely by defect accumulation, but rather by orientation-sensitive interactions among dislocations, defects, and fracture process zones.
A key novelty of this work lies in integrating radiation-induced defect evolution with orientation-dependent fracture within an atomistic T–S analysis, enabling quantitative assessment of atomic-scale fracture resistance under realistic defect conditions.
\footnotetext[2]{Published in \emph{International Journal of Plasticity} \textbf{201}, 104686, 2026, doi: 10.1016/j.ijplas.2026.104686}
\end{abstract}


\begin{keyword}
Crack propagation \sep Radiation defects \sep MD simulations \sep Cr-rich alloy \sep T--S law \sep Atomic-scale fracture energy 
\end{keyword}

\end{frontmatter}

\input{sections/1_Introduction}

\input{sections/2_Simulation_models}

\input{sections/3_Physical_mechanisms}

\input{sections/4_Mechanical_responses}

\input{sections/5_Summary}

\section*{Acknowledgments}
The National Science Centre (NCN) in Poland supports this work through Grants No.\ 2020/38/ E/ST8/00453 (HM and AU) and 2022/47/I/ST8/02879 (SS). 
We gratefully acknowledge Polish high-performance computing infrastructure PLGrid (HPC Center at ACK Cyfronet AGH) for providing computer facilities and support within computational Grant No.\ PLG/2024/017084.
We also thank F.J.\ Dominguez-Gutierrez for carrying out the collision cascade simulations.

\appendix
\renewcommand{\thefigure}{A.\arabic{figure}}
\setcounter{figure}{0}
\section{Orientation-based evaluation of crack-tip stress distributions}
\label{sec:appendixA}
The supplementary stress-field maps in this appendix provide additional insight into the orientation-dependent mechanisms discussed in Sections~\ref{sec:Section3} and~\ref{sec: Section4}.

\begin{figure}[h]
    \centering    \includegraphics[width=0.8\textwidth]{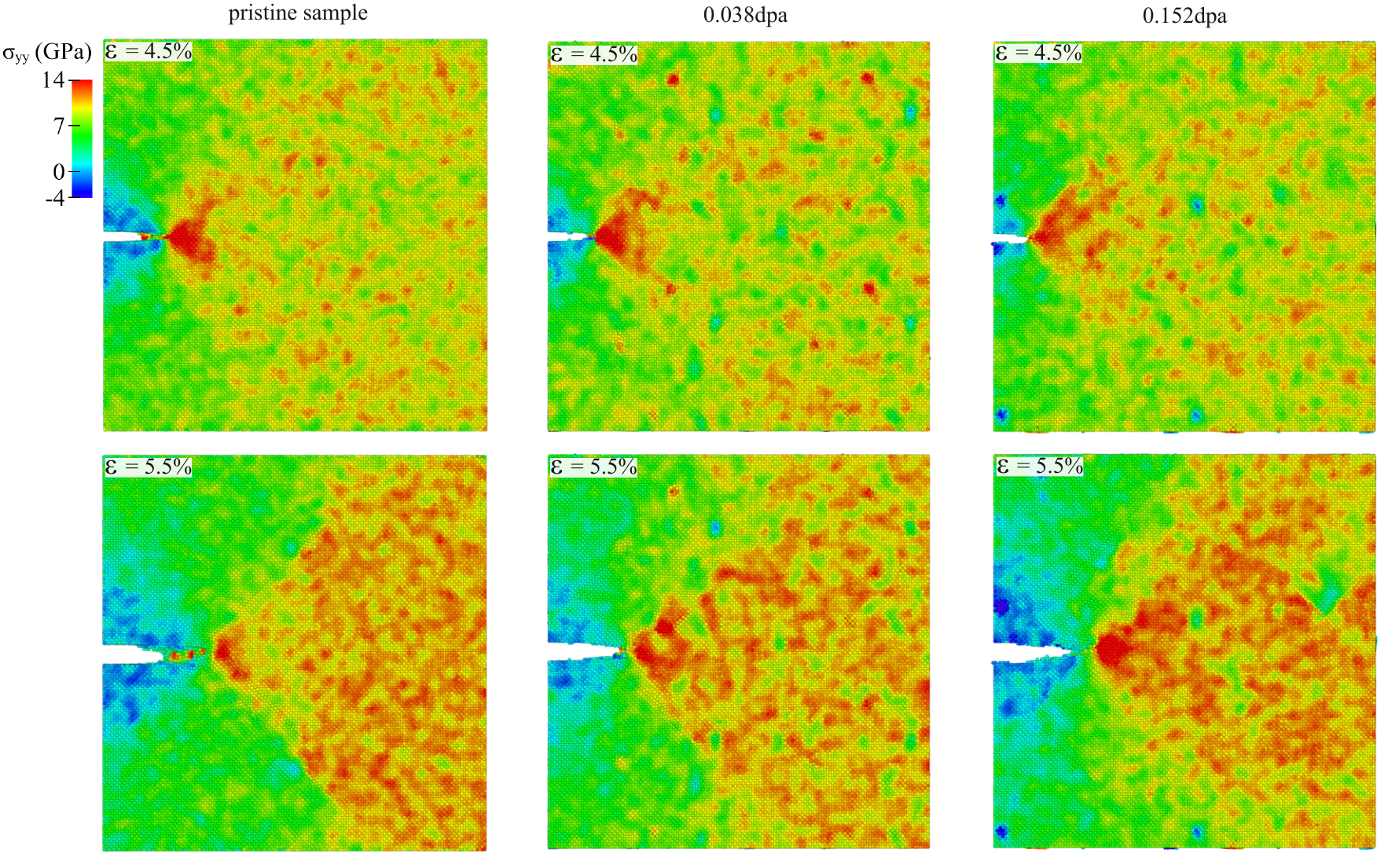}
    \caption{
Stress field distribution ($\sigma_{yy}$) in pristine and irradiated samples with the (001) crystallographic orientation during crack propagation, corresponding to the $\{010\}\langle001\rangle$ crack configuration.}
\phantomsection
\label{fig:stress(001)}
\end{figure}

\begin{figure}[h!]
    \centering
    \includegraphics[width=0.8\textwidth]{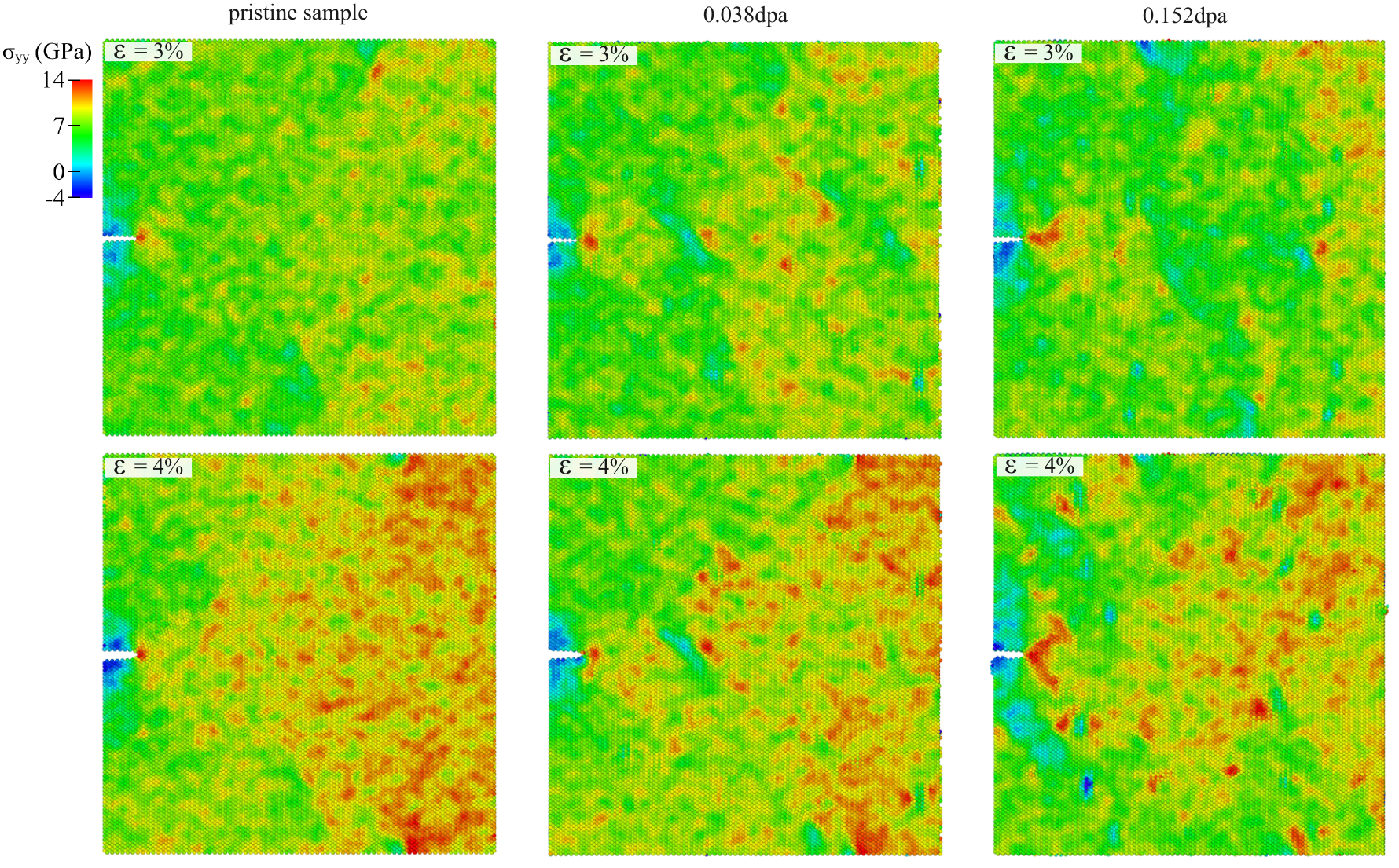}
    \caption{Stress field distribution ($\sigma_{yy}$) in pristine and irradiated samples with the (011) crystallographic orientation during crack propagation, corresponding to the $\{01\bar{1}\}\langle011\rangle$ crack configuration.}
\phantomsection
\label{fig:stress(011)}
\end{figure}

\begin{figure}[h!]
    \centering
    \includegraphics[width=0.8\textwidth]{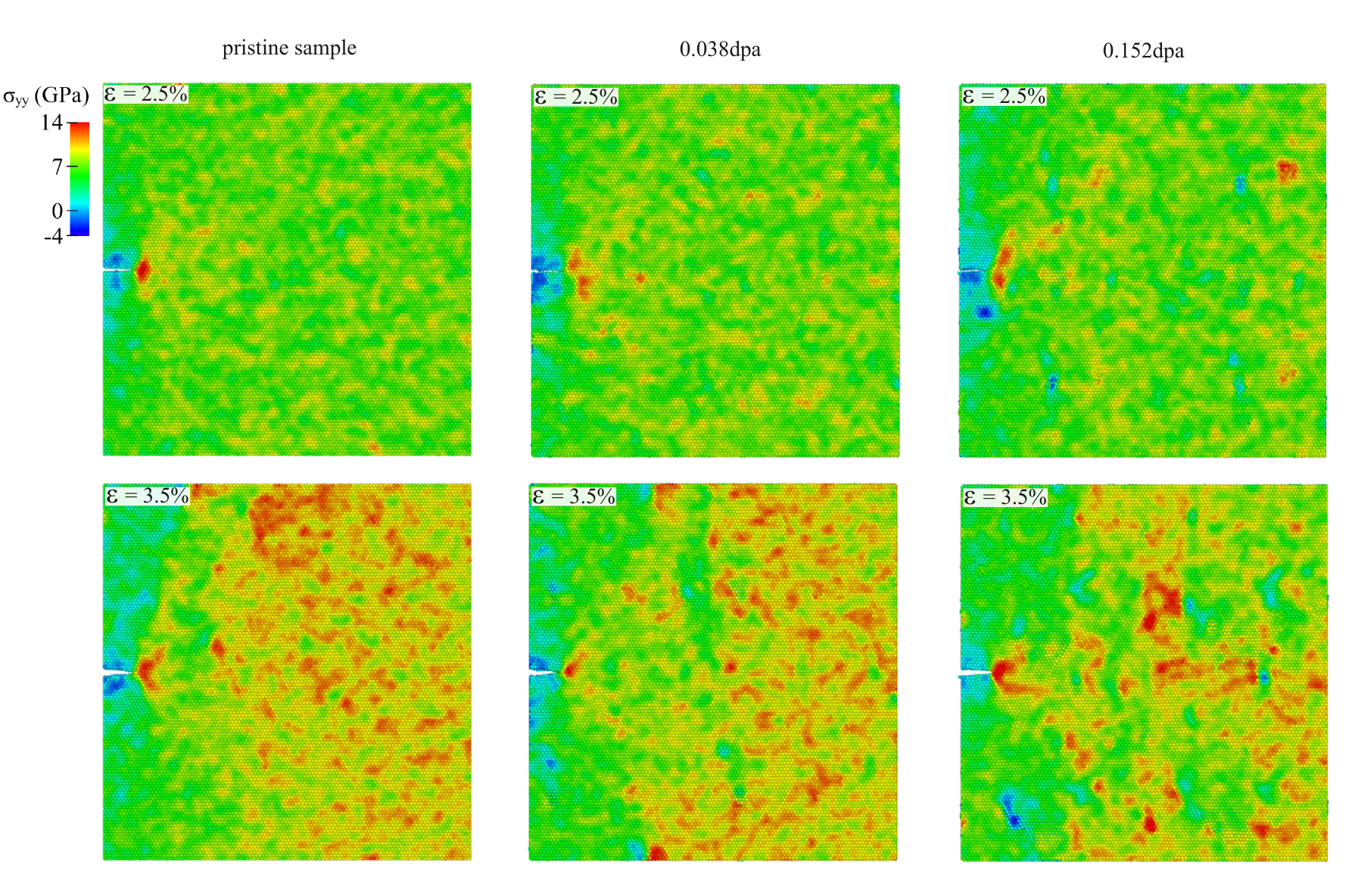}
    \caption{Stress field distribution ($\sigma_{yy}$) in pristine and irradiated samples with the (111) crystallographic orientation during crack propagation, corresponding to the $\{\bar{1}01\}\langle111\rangle$ crack configuration.}
\phantomsection
\label{fig:stress(111)}
\end{figure}

Each representative cross-section presents the crack-tip $\sigma_{yy}$ distributions for three irradiation levels, including pristine sample, 0.038~dpa, and 0.152~dpa, evaluated at two strain levels specific to each crystallographic orientation. For the (001) orientation, the stress fields are shown at $\varepsilon = 4.5\%$ and $5.5\%$ in Fig.~\ref{fig:stress(001)}; for the (011) orientation at $\varepsilon = 3.0\%$ and $4.0\%$ in Fig.~\ref{fig:stress(011)}; and for the (111) orientation at $\varepsilon = 2.5\%$ and $3.5\%$ in Fig.~\ref{fig:stress(111)}. These results illustrate how irradiation alters the spatial distribution of $\sigma_{yy}$ for the three crystallographic orientations and their corresponding crack configurations. To enhance the clarity of the crack-tip stress contours, the $\sigma_{yy}$ fields shown in these results were post-processed using a neighbor-mean stress calculation, in which virial stresses were smoothed over a 5~\AA{} radial neighborhood to reduce local noise and highlight the underlying spatial trends.

For the (001) orientation, the stress fields in Fig.~\ref{fig:stress(001)} reveal a consistently sharp and highly localized tensile peak positioned directly ahead of the crack tip. With increased irradiation dose, this peak intensifies and becomes progressively narrower. The minimal lateral spreading of stress contours highlights the inherently limited plastic relaxation capacity of the (001) system. Irradiation amplifies this limited plasticity by further constraining dislocation mobility, in agreement with the mechanisms discussed in Sections~\ref{sec:Section3.2} and~\ref{sec: Section4.1}, where slip--defect interactions were shown to suppress crack-tip plasticity and produce the pronounced hardening and reduced ductility observed in the stress–strain responses.

The (011) orientation exhibits the most pronounced radiation-induced transformation among the three, as shown in Fig.~\ref{fig:stress(011)}. While the pristine configuration already displays a relatively confined tensile zone, higher irradiation levels significantly sharpen the crack-tip stress peak and generate asymmetric secondary hotspots. These features indicate strong obstruction of slip by radiation-induced defects, leading to severe local stress accumulation and a highly constrained plastic zone. This response is fully consistent with the behavior reported in Section~\ref{sec:Section3.3}, where the (011) orientation was shown to be particularly susceptible to defect clustering and slip inhibition. Moreover, the strong localization of stress correlates directly with the steep hardening and reduced plastic flow observed in the stress--strain curves of Section~\ref{sec: Section4.1}, as well as with the accelerated crack advance and higher crack velocity exhibited by irradiated (011) samples.

In contrast, the (111) orientation preserves the broadest and most diffuse crack-tip stress field, as illustrated in Fig.~\ref{fig:stress(111)}. Even with increasing irradiation levels, the tensile hotspot sharpens only slightly, and the stress contours continue to extend laterally, indicating that the plastic zone remains comparatively large and well developed. This behavior reflects the high availability of favorably oriented slip systems on the (111) plane, which promotes efficient stress redistribution and enables dislocations to relax the crack tip, consistent with the slip-activation mechanisms discussed in Section~\ref{sec:Section3.4}. The smooth stress gradients and lack of significant secondary hotspots also align with the mechanical response described in Section~\ref{sec: Section4.1}, where the (111) samples exhibited the mildest hardening rate and the most stable stress–strain behavior.

A comparison of the three orientations highlights a clear hierarchy in irradiation sensitivity, stress localization, and DBT behavior. The (001) orientation exhibits inherently limited plasticity, with stress localization that intensifies under irradiation and promotes an early shift toward brittle-like crack advance. The (011) orientation exhibits the most severe irradiation-driven stress concentration, developing sharply focused tensile peaks and secondary hotspots that reflect strong defect–crack interactions and accelerate the onset of DBT. In contrast, the (111) orientation maintains the broadest tensile stress distribution and the most effective plastic relaxation even at elevated dpa levels, delaying the DBT and preserving a more ductile response under irradiation.

\bibliographystyle{elsarticle-harv}
\bibliography{references}

\end{document}

%% file: sections/1_Introduction.tex
\section{Introduction}
\label{sec:Section1}

Materials exposed to high-energy radiation, such as neutron irradiation, undergo profound microstructural transformations that critically impact their mechanical properties. 
In nuclear structural environments, including high-temperature components in both fission reactors and fusion systems located behind plasma- facing layers (e.g., blanket structural supports and shield blocks), austenitic alloys are frequently employed due to their high-temperature strength, corrosion resistance, and ductility. Among these, 310S stainless steel is particularly relevant as a candidate for radiation-exposed structural elements, because its high chromium and nickel content confer excellent oxidation resistance combined with good creep stability \citep{Ortner2023}.

These alloys are valued for their swelling resistance, good irradiation creep behavior, and relative ease of fabrication \citep{LUCON2006, STORK2014}. The high-temperature strength and low carbon content of 310S make it a suitable model alloy for investigating how irradiation influences defect formation, hardening, and embrittlement under service-relevant conditions. However, radiation-driven changes significantly affect mechanical integrity and can lead to substantial material degradation.

This radiation-induced degradation manifests as embrittlement, hardening, and reduced ductility, posing serious challenges to the structural integrity of materials operating in extreme environments, including both nuclear fission and nuclear fusion reactors \citep{zinkle2012, Nordlund2018_dpa}.
Understanding these radiation-induced effects is crucial for designing materials with controlled defect evolution in radiation-intensive environments.

In irradiated crystalline materials, the microstructural response to deformation also depends on crystallographic orientation, which governs the migration of radiation-induced defects and their interaction with other microstructural features, such as dislocations and grain boundaries, that influence mechanical behavior. The radiation-induced defects influence slip systems activity, localized plastic deformation, and fracture behavior, thus contributing to anisotropy in mechanical properties \citep{CUI2017, TSUGAWA2022, BAO2022, Ustrzycka2024, XIA2025}.
The same defect structures trigger orientation-selective deformation pathways. \citet{CUI2017} and \citet{TSUGAWA2022} demonstrate that dislocations loops immobilize specific slip systems when their Burgers vectors align with preferred defect-migration directions, whereas \citet{BAO2022} and \citet{Ustrzycka2024} show that dislocations can bypass obstacles through cross-slip or defect-assisted channeling depending on lattice orientation. 
\citet{XIA2025} shows that this orientation-dependent interplay between defects and slip determines whether strain localizes or remains distributed, establishing crystallographic orientation as a factor controlling anisotropic post-irradiation mechanical response.

The orientation of mechanical loading with respect to the crystal lattice also influences the fracture mechanics of irradiated materials by altering crack propagation pathways, strain localization, evolution of plastic mechanisms, and the ductile-to-brittle transition (DBT) behavior \citep{klueh1997neutron, BARIK2023101667, LIN2025, Ustrzycka2025}. Orientations with a high density of active slip systems can promote the activation of slip systems that enhance stress redistribution, while orientations with fewer active slip systems can lead to localized stress concentrations and accelerated crack growth \citep{chen2022, XIE2023, JIAN2024, LAI2025}. Understanding the orientation-dependent mechanisms and anisotropic effects is crucial in determining the mechanical performance of irradiated materials \citep{WANG2020, LI2022, HUANG2024}.

The extent to which radiation-induced defects modify anisotropic behavior, particularly in relation to defect evolution, strain localization, and fracture mechanisms, is an important aspect considered in this work.

To capture the influence of orientation-dependent, defect-driven plasticity on crack initiation and propagation, an atomistically-informed continuum framework can provide a suitable approach. Fracture mechanics offers a robust framework for analyzing deformation and failure in irradiated materials. Stress-based parameters, such as the stress intensity factor (SIF) \citep{WILSON2019, ZHAO2025, FU2025}, provide a measure of crack-tip stresses but cannot capture the nanoscale, defect-driven energy dissipation that dominates crack-tip behavior in molecular dynamics simulations. Energy-based methods, including the $J$-integral \citep{KIM20211, JIA2022} and the critical strain energy release rate ($G_{\rm c}$) \citep{BARIK2023101667, LIN2025, Ustrzycka2025}, determine bulk fracture energetics but assume continuous, homogeneous plasticity and small-scale yielding. 
When the plastic zone is large, such that it occupies the whole computational domain, the $J$-integral approach is not applicable. Likewise it is not applicable in the case of irradiated materials where the defects are distributed within the whole volume and inelastic deformations initiate at those defects.

Traction--separation (T--S) laws \citep{ZENG2018120, wang2021molecular, Xie2021, YU2024} explicitly relate crack-tip traction to atomistic deformation mechanisms, providing a physically grounded approach to probe anisotropic fracture under irradiation. These complementary methods are widely employed in both continuum and atomistic simulations \citep{CHOWDHURY2016}.

In irradiated materials, fracture behavior is governed by DBT. Radiation-induced DBT, driven by defect accumulation, nano-voids absorption, and dislocation pinning, fundamentally differs from the extensively studied temperature-driven DBT \citep{BARIK2023101667, Ustrzycka2025}. 
In particular, radiation-induced DBT can occur at temperatures where the pristine material remains ductile, because irradiation-induced defects suppress dislocation-mediated plasticity and govern the fracture response.
Moreover, DBT is expected to be sensitive to crystallographic orientation, as slip system activity and stress redistribution vary with lattice directionality. Additionally, the anisotropic nature of crack propagation influences the interaction between radiation-induced defects, dislocations, and crack fronts, potentially accelerating or delaying DBT depending on the available deformation mechanisms. Our previous work on the (001) orientation \citep{Ustrzycka2025} provided an initial evidence of this distinction. In the present study, we broaden this perspective by systematically analyzing the influence of crystallographic orientation on DBT mechanisms and by applying a T--S approach instead of the energy-based characterization of fracture energy used in \citet{Ustrzycka2025}.

One of the aims of this study is to determine how crystallographic orientation modulates DBT in irradiated materials. This aim reflects the need to establish how orientation-dependent deformation pathways translate into different crack-tip responses under irradiation, where fracture is controlled by the competition between defect-limited plasticity and crack advance.
Elucidating this link at the atomistic scale contributes to a mechanistic understanding of irradiation-induced intragranular fracture in polycrystalline aggregates, where local crystal orientation governs whether deformation ahead of a crack remains accommodating or becomes brittle.

Molecular dynamics (MD) simulations provide an indispensable tool for comprehensively studying these complex behaviors at the nanoscale \citep{Zhou2003, Osetsky2004, stukowski2010extracting}. By capturing the atomic-scale interactions between radiation-induced defects, dislocations, and crack propagation mechanisms, MD simulations enable the detailed investigation of microstructural evolution and mechanical responses that are otherwise inaccessible through experimental techniques alone. These simulations allow for a precise control of crystallographic orientation, irradiation conditions, and mechanical loading conditions, offering unique insights into the anisotropic effects of radiation on material properties \citep{shimokawa2005defect,yin2007new, chen2022}. Furthermore, MD simulations facilitate the exploration of defect dynamics, such as the formation and migration of vacancies, self-interstitial atoms (SIAs), and dislocation loops, as well as their influence on fracture mechanics \citep{bitzek2008atomistic, WANG2015, CUI2017, TSUGAWA2022, BAO2022, LIN2024, Ustrzycka2024}. This capability is particularly valuable for understanding the DBT behavior and the role of slip systems in stress redistribution \citep{barrows2016traction, BARIK2023101667}. 

This study employs the T--S law within MD simulations to systematically investigate the anisotropic effects of crystallographic orientation on crack propagation and its influence on the DBT in irradiated alloys to extract the atomic-scale fracture energy under realistic defect conditions.
By analyzing crack initiation and growth across three high-symmetry orientations, this study provides a comprehensive assessment of how crystallographic orientation influences anisotropic fracture behavior and DBT. 

We identify distinct deformation mechanisms across orientations, ranging from defect-driven blocking of stacking faults and dislocation pinning in the (011), to stable plastic accommodation in the (111), and constrained slip in the (001), and link them directly to quantitative changes in atomic-scale fracture energy. These orientation-dependent mechanisms contribute to the development and enhancement of mechanical anisotropy in irradiated materials by distinctly influencing defect evolution, slip system activation, and strain localization, which lead to significant directional differences in fracture resistance and energy dissipation.
This work thus provides a physics-grounded understanding of radiation-induced fracture processes and defines the role of crystallographic orientation in governing post-irradiation mechanical performance.

This study provides a comprehensive atomistic perspective, based on a single-crystal model, on radiation-induced embrittlement in an fcc $\rm Fe_{55}Ni_{19}Cr_{26}$ alloy, revealing how crystallographic orientation mediates fracture behavior through its influence on plasticity and defect interactions. The defect-driven 
DBT is investigated by keeping the temperature constant in the conditions in which the pristine material remains fully ductile. The intrinsic effect of radiation-induced defects on crack-tip deformation and fracture is thus isolated. The novelty of this work lies in establishing an atomistic framework that couples radiation-induced defect evolution with crystallographic-orientation–dependent fracture. By integrating T–S analysis directly within MD simulations, the approach allows evaluation of atomic-scale fracture energy under realistic defect conditions. 
This approach emphasizes that radiation-induced embrittlement reflects the interplay between defect–interaction mechanisms and the accessibility of specific plasticity modes, offering a framework that can be applied to explore orientation-sensitive embrittlement phenomena across other crystallographic configurations. To the best of our knowledge, no published study has addressed radiation-driven DBT within a unified atomistic framework, and we believe that the present work provides a substantive contribution to the field.

The paper is organized as follows: Section \ref{sec:Section2} describes the MD simulation methodology for a single-crystal model, including the preparation of irradiated alloy samples for different crystallographic orientations and the computational setup for tensile crack propagation analysis.
Section \ref{sec:Section3} examines the anisotropic effects of radiation-induced defects on fracture behavior, focusing on slip system activation, dislocation-defect interactions, and deformation response near the crack tip. Section \ref{sec: Section4} presents the T--S law framework to quantify the atomic-scale fracture energy and assess DBT behavior across different orientations.

%% file: sections/2_Simulation_models.tex
\section{Simulation models}
\label{sec:Section2}

This study involves a two-stage MD simulation approach. Subsection~\ref{sec:Section2.1} details the methodology for generating radiation damage in material samples via overlapping collision cascade simulations. The resulting pre-irradiated configurations serve as input for the next stage. Subsection~\ref{sec:Section2.2} presents the setup and preparation of numerical models for tensile crack propagation tests on these pre-irradiated samples.

\subsection{Generation of irradiated samples via overlapping collision cascades}
\label{sec:Section2.1}
\vspace{0.2cm}

The material modeled in this work is an fcc $\rm Fe_{55}Ni_{19}Cr_{26}$ alloy. The adopted alloy composition corresponds to the 310S stainless steel. The initial atomic configurations were constructed using the Atomsk tool \citep{HIREL2015}, aligning the simulation boxes along three principal crystallographic orientations: (001), (011), and (111), as summarized in Table~\ref{Tab:orientation}. These orientations were selected to capture the representative behavior of high-symmetry directions in fcc lattices, providing distinct combinations of active slip systems and allowing evaluation of how crystallographic geometry influences plasticity, defect interactions, and crack propagation under irradiation.
The resulting configurations then serve as the starting point for irradiation simulations using MD, with the detailed procedure for alloy generation, lattice optimization, and relaxation described in \citet{Ustrzycka2024, Ustrzycka2025}.

Radiation-induced defect formation was simulated using MD with a sequential overlapping collision cascade method, following the approach in \citet{Nordlund2018_dpa, Toussaint2021, ZHOU2023, Ustrzycka2024}. 
The method involves a sequence of collision-cascade simulations carried out within the same simulation cell, where defects produced by earlier cascades persist and can interact with those generated by subsequent cascades \citep{Nordlund2018_dpa}. This methodology has been previously validated for Fe--Cr--Ni alloys by comparison with transmission electron microscopy (TEM) observations, showing good quantitative agreement in predicted defect densities \citep{Ustrzycka2024}.

In each cascade, a single primary knock-on atom (PKA) was generated by imparting 10~keV of kinetic energy to a randomly selected Fe, Ni, or Cr atom located within the central region of the simulation box. The velocity corresponding to this energy was computed using the Fe atomic mass and assigned in a randomly chosen direction. Using the same velocity for all PKAs leads to only a small deviation from the nominal 10~keV energy in view of the similar atomic masses of Fe, Ni, and Cr, and this deviation is not expected to significantly affect the overall defect production or cascade morphology.

In service conditions, PKAs are produced by neutrons from fission reactions, which are born with energies on the order of ~0.1–2 MeV in typical light‑water reactors \citep{Maslov2010, TRKOV2015}. This generates PKAs with energies spanning from a few keV up to several hundred keV, so the energy of the PKA used in the simulations can be considered broadly representative of recoil events relevant to the early stages of defect formation in austenitic steels \citep{GILBERT2015}. The defect production depends on the PKA energy, with lower-energy PKAs (1–10~keV) generating mostly local collision cascades with small vacancy and self-interstitial clusters, while higher-energy PKAs around 100~keV or more tend to produce larger cascades with bigger defect clusters and subcascades \citep{Nordlund2018_dpa}. Note, however, that the PKA energy employed in the MD simulations is limited by the simulation cell size so that interaction of a cascade with its periodic image is avoided.

In total, 400 sequential overlapping cascades were simulated for each crystallographic orientation, producing an accumulated dose of 0.152~dpa (displacements per atom, i.e., the average number of times each atom in the material is displaced from its lattice site due to irradiation \citep{Nordlund2018_dpa}). Collision cascade simulations were performed at 300~K with periodic boundary conditions applied in all directions, and the box size (see Table~\ref{Tab:orientation}) was chosen sufficiently large to prevent interaction between cascades and their periodic images. This setup guarantees the physical independence of each cascade while maintaining computational efficiency. A detailed analysis of the simulation box dimensions for various PKA energies in $\rm Fe_{55}Ni_{19}Cr_{26}$ alloys is provided in \citet{Ustrzycka2024}.

MD simulations were performed using the LAMMPS package \citep{LAMMPS}, with an interatomic potential based on the Embedded Atom Method (EAM), specifically parameterized for Fe--Ni--Cr alloys (EAM11) \citep{Bonny2011}. 
This potential includes a built-in short-range ZBL-type correction for high-energy atomic collisions and provides a reliable representation of atomic interactions and defect formation processes, making it well-suited for modeling radiation effects in fcc alloys \citep{BELAND2017, ZHOU2023, Ustrzycka2024}.
The simulations were carried out in the NVT ensemble, with the temperature control applied using a Langevin thermostat. Each cascade was run sequentially for 15 000 steps using an adaptive timestep capped at 0.1 fs, ensuring post-ballistic relaxation and thermal equilibration prior to the subsequent event. After completion of the cascade run, the system was equilibrated at 300~K to allow stabilization of the generated defects before initiating the next cascade. 
Moreover, electronic stopping, acting as a frictional force on the PKA, has not been included in the MD simulations. For low PKA energies in metals (around 10 keV), its effect on defect production is minor compared to predictions from the conventional Norgett--Robinson--Torrens (NRT) model \citep{Nordlund2018_dpa, Ustrzycka2024}.

\begin{table}
\centering
\caption{Specification of the simulation cells used in the overlapping collision cascade simulations.}

\vspace{2ex}

\renewcommand{\arraystretch}{1.2} 
{\small
\begin{tabular}{*{6}{c}} 
\hline 
Orientation & $x$-axis & $y$-axis & $z$-axis & Dimensions (\text{\AA}\(^3\)) & Number of atoms \\ 
\hline 
(001) & [100] & [010] & [001] & 138 × 141 × 148 & 262080 \\ 
(011) & [100] & [01$\bar{1}$] & [011] & 138 × 142 × 147 & 262314 \\ 
(111) & [1$\bar{2}$1] & [$\bar{1}$01] & [111] & 137 × 142 × 147 & 261360 \\ 
\hline
\label{Tab:orientation}
\end{tabular}
}
\footnotesize
\end{table}

\begin{figure}[b!]
    \centering
 \includegraphics[scale=0.41]{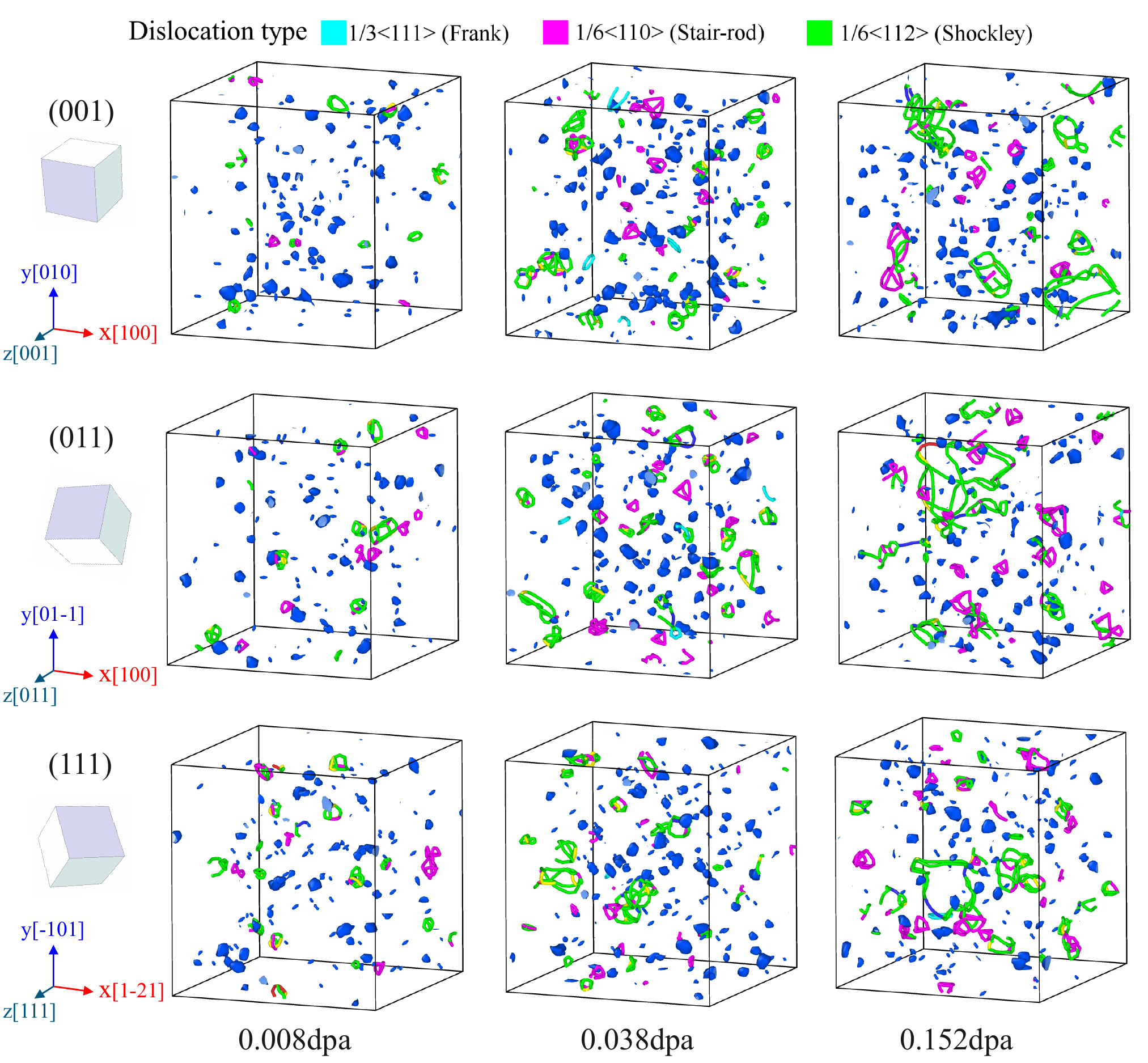}
    \caption{Fe$_{55}$Ni$_{19}$Cr$_{26}$ samples irradiated using sequential collision cascades at three damage levels (0.008, 0.038, and 0.152 dpa) for the crystallographic orientations (001), (011), and (111). The legend indicates the types of dislocation loops. Voids are marked as blue spheres.}
    \label{fig:orientations}
\end{figure}

\begin{figure}[t!]
    \centering
   \includegraphics[scale=0.7]{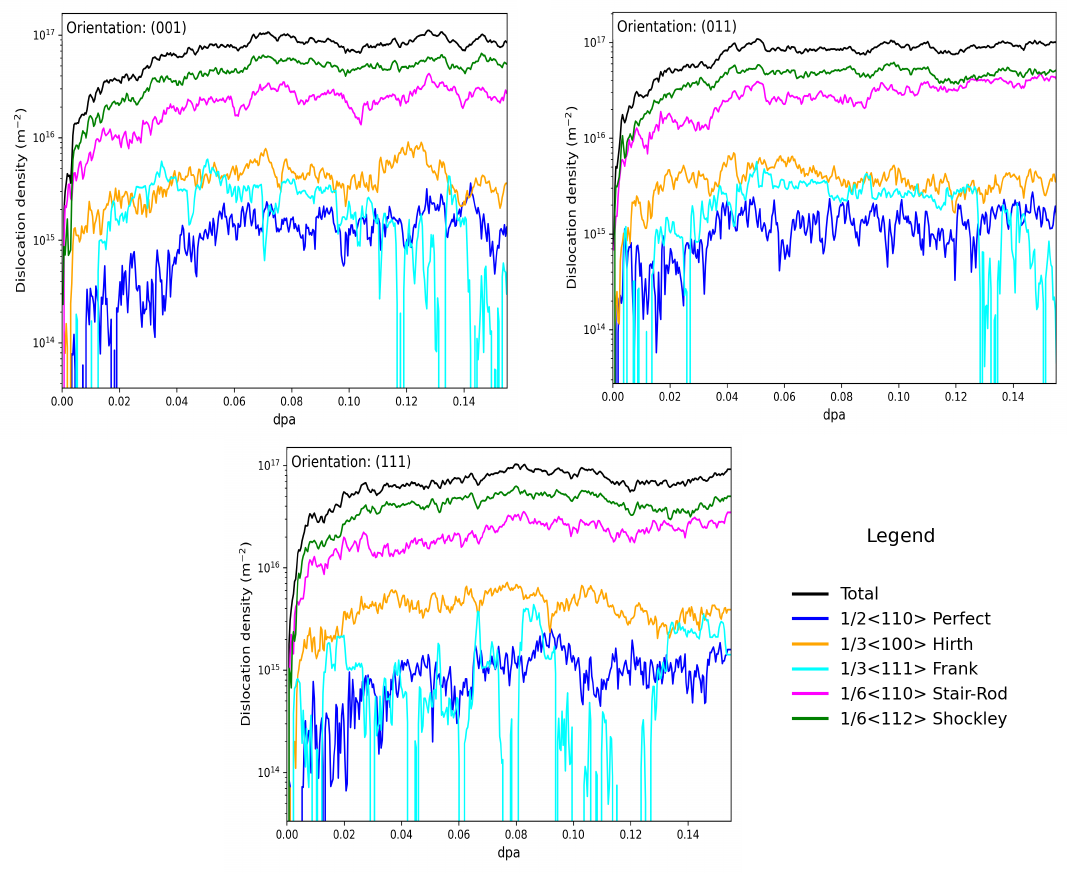}
       \caption{Evolution of dislocation density as functions of irradiation dose (dpa) for the (001), (011) and (111) orientations.}
    \label{fig:dislocations_density}
\end{figure}
\begin{figure}[b!]
    \centering
   \includegraphics[scale=0.5]{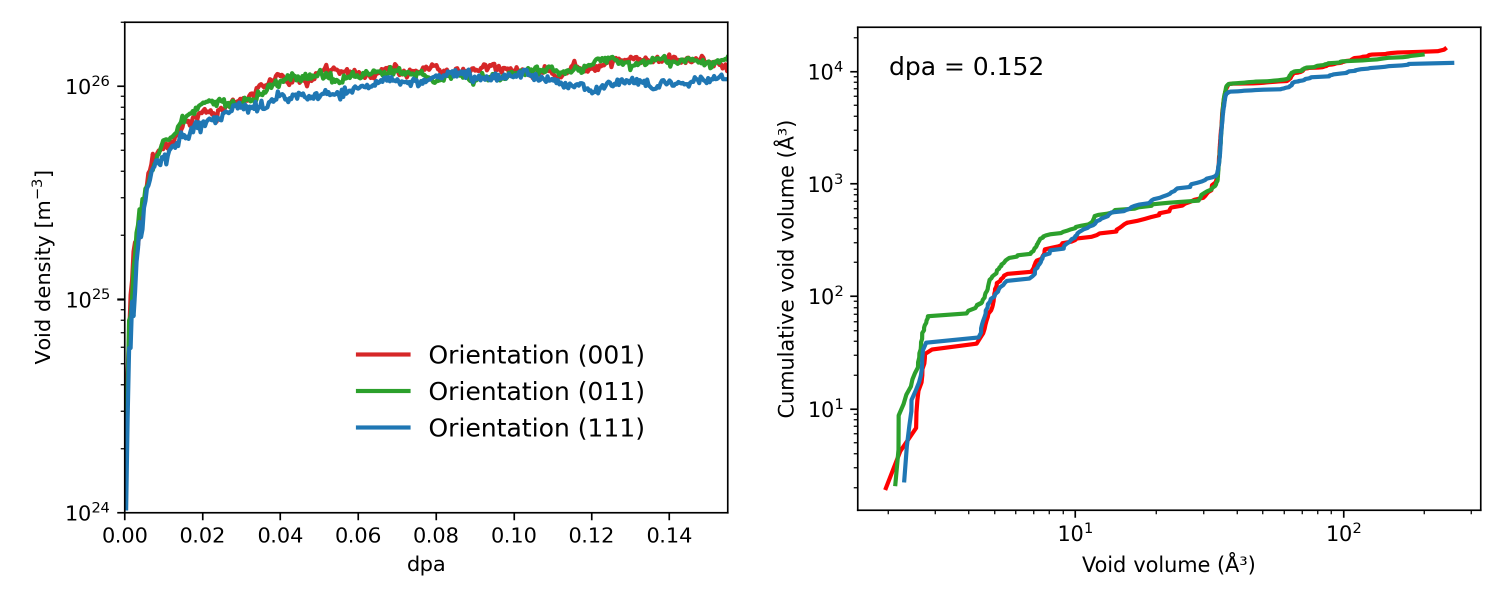}
       \caption{Evolution of void number density (left) and cumulative void volume distribution (right) as functions of irradiation dose for the (001), (011), and (111) orientations.}
    \label{fig:voids_density}
\end{figure}
\begin{figure}[t!]
    \centering
   \includegraphics[scale=0.22]{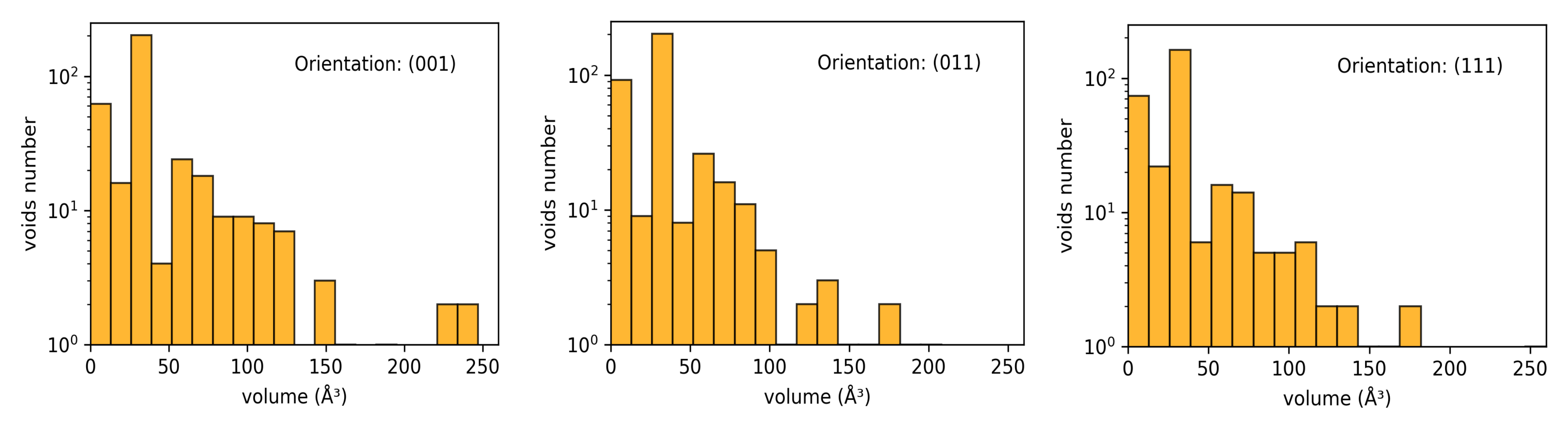}
   \vspace{-3ex}
       \caption{Populations of radiation-induced voids at dpa~=~0.152 for selected orientations.}
    \label{fig:histogram}
\end{figure}
The study considers samples irradiated to 0.008, 0.038, 0.076 and 0.152 dpa (20, 100, 200 and 400 cascades, respectively), which were selected to cover a range from very low to moderate damage levels, representative of the early to intermediate stages of irradiation in austenitic steels under typical fission reactor conditions. Fig.~\ref{fig:orientations} presents the defect structures in selected irradiated samples.
For the analysis of radiation-induced defects in the MD simulation results, the Dislocation Extraction Algorithm (DXA) and Common Neighbor Analysis (CNA) tools implemented in the OVITO software were employed \citep{ovito}. These methods enable detailed visualization and quantification of microstructural features in irradiated materials.

Primary radiation events produce only vacancies and interstitial atoms (compare \citet[\mbox{Fig.~2}]{Ustrzycka2024}). During subsequent collisions, radiation defects evolve and grow; vacancies aggregate into clusters that stabilise into voids, represented in Fig.~\ref{fig:orientations} as blue spheres. Helium bubbles, which can form in materials through (n, $\alpha$) transmutation reactions depending on neutron energy and fluence, are not generated and analyzed in this study. In fission reactor conditions, helium production in 310S-type austenitic steels is low and accumulates slowly, so its effect on radiation-induced embrittlement is minor. In structural regions of fusion systems where neutron fluence is substantially reduced compared to first-wall zones, helium accumulation is also limited and does not dominate defect-driven fracture mechanisms under the present conditions. Other radiation-induced defect structures play a key role in fcc metals. In particular, the relatively low stacking fault energy facilitates the formation of stacking fault tetrahedra (SFTs) \citep{Nordlund1999}, which adopt a pyramidal morphology and are visible in Fig.~\ref{fig:orientations} as pink-colored pyramidal defects.
Self-interstitial atoms, in turn, form clusters that collapse into dislocation loops \citep{HERNANDEZMAYORAL2010, Ustrzycka2021}. With continued irradiation, these loops interact, evolve, or annihilate. 

Radiation-induced defects in fcc metals exhibit nearly orientation-independent characteristics, primarily due to the weak directional dependence of the threshold displacement energy (TDE), which remains approximately constant in austenitic Fe–Ni-Cr alloys (\textasciitilde\!40 eV \citep{Nordlund2018_dpa}). Moreover, fcc lattices lack pronounced channeling paths, dominant in more open bcc structures, which strongly reduces the possibility of directionally enhanced defect transport \citep{MAHROUS2025}. The dense atomic packing in fcc alloys promotes rapid cascade branching, which suppresses directional transport of defects. As a result, defect production is largely independent of lattice orientation.

The qualitative defect morphologies shown in Fig.~\ref{fig:orientations} are complemented by a quantitative analysis in Figs.~\ref{fig:dislocations_density}--\ref{fig:histogram}. 
Fig.~\ref{fig:dislocations_density} presents the evolution of the total and resolved dislocation densities as a function of the irradiation dose. 
The results show a rapid nucleation of dislocation loops and voids at low doses followed by growth and partial saturation of defect densities as the dose increases, a behavior observed in atomistic cascade studies \citep{Nordlund2018_dpa, BELAND2017, ZHOU2023}. 
The kinetics is very similar for the (001), (011), and (111) orientations, which proves that the defect formation is not affected by the orientation of the MD sample. The apparent fluctuations of Frank dislocation density, particularly in the (111) orientation, are a consequence of the logarithmic scale used for the vertical axis, while the actual fluctuations are small.

Fig.~\ref{fig:voids_density} presents the corresponding evolution of voids. The left panel shows the void number density, while the right panel shows the cumulative void volume distribution, obtained by sorting voids by size and computing the cumulative sum of their volumes. The latter reveals that all orientations generate void populations with similar cumulative growth, indicating comparable rates of vacancy clustering.

Finally, Fig.~\ref{fig:histogram} displays a histogram of the void-size distributions at 0.152~dpa. All three orientations show similarly shaped distributions dominated by small voids, with only minor variations in the largest voids.

Given the observed defect distribution independent of MD sample orientation (Fig.~\ref{fig:orientations}) and comparable evolution of defect densities and void volume fractions across the investigated orientations (Figs.~\ref{fig:orientations}--\ref{fig:histogram}), the current MD-based framework provides a sufficiently robust representation of radiation-induced defect populations. Therefore, no additional formal statistical analysis of individual collision cascades was considered necessary.

\subsection{Setup for orientation-dependent tensile crack simulation}
\label{sec:Section2.2}
\vspace{0.2cm}
A crack propagation model was developed to study how the mechanical loading affects crack growth behavior for different crack orientations. The MD model with an initial pre-crack under tensile loading is shown in Fig. \ref{fig: Mode-i}. 

The initial irradiated samples, with three crystallographic orientations as shown in Fig.\ref{fig:orientations}, were replicated to create three distinct crack growth models. The replication method involved doubling the simulated sample dimensions in the $x$- and $y$-directions, while the $z$-direction remained unchanged. Ultimately, considering the dimensions of the initial irradiated samples, the model with approximate dimensions of $275\times282\times148$~\text{\AA}\(^3\) and containing more than 1 million atoms (varying slightly for different orientations due to differing lattice structures) was established. 

Fig.~\ref{fig: Mode-i} comprises two views to illustrate the three-dimensional simulated sample fully. The left side of Fig.~\ref{fig: Mode-i} shows the $x$--$y$ projection of the pre-cracked sample. The right side presents an $y$-$z$ projection, emphasizing the model's depth along the $z$-direction. {Table~\ref{Tab:crack_models_replicated} summarizes the parameters of the replicated crack models, including the crack geometry using the $\{\text{crack plane}\}\langle\text{crack-front direction}\rangle$ notation, loading axis, maximum Schmid factors for slip ($m^{\rm slip}_{\rm max}$) and twinning ($m^{\rm twin}_{\rm max}$), and cell dimensions.
These three crack configurations were chosen to represent crystallographic planes and crack-front directions that probe distinct fracture responses in fcc materials. In particular, they involve different geometric relationships between the crack plane, crack front, and the primary $\{111\}\langle1\bar{1}0\rangle$ slip systems, which are known to control crack-tip plasticity and dislocation emission.

For all three crystallographic orientations, the maximum Schmid factor for dislocation slip, $m^{\rm slip}_{\rm max}$, attains the same value. This enforces crystallographic equivalence of the primary $\{111\}\langle1\bar{1}0\rangle$ slip systems with respect to the applied uniaxial tension along the sample's $y$-direction, providing a consistent framework for comparing the selected crack geometries. In contrast, the maximum Schmid factor for deformation twinning, $m^{\rm twin}_{\rm max}$, varies with orientation, with the (001) configuration exhibiting a lower value than the (011) and (111) cases, reflecting the differing alignment of the $\{111\}\langle11\bar{2}\rangle$ twinning systems relative to the load. 
At the same time, only for the (011) orientation, the nominal crack front (parallel to the $z$-axis) is orthogonal to the shear plane of (two) twinning systems.
While fcc steels (except for the high-manganese TWIP steels) plastically deform predominantly via dislocation slip under conventional loading, the highly heterogeneous stress state and local instabilities at the crack tip can nucleate twin-like stacking arrangements, see e.g.\ \cite{ANDRIC2018144}, as indeed observed in some simulations reported below.

\begin{figure}[b!]
    \centering
\includegraphics[scale=0.38]{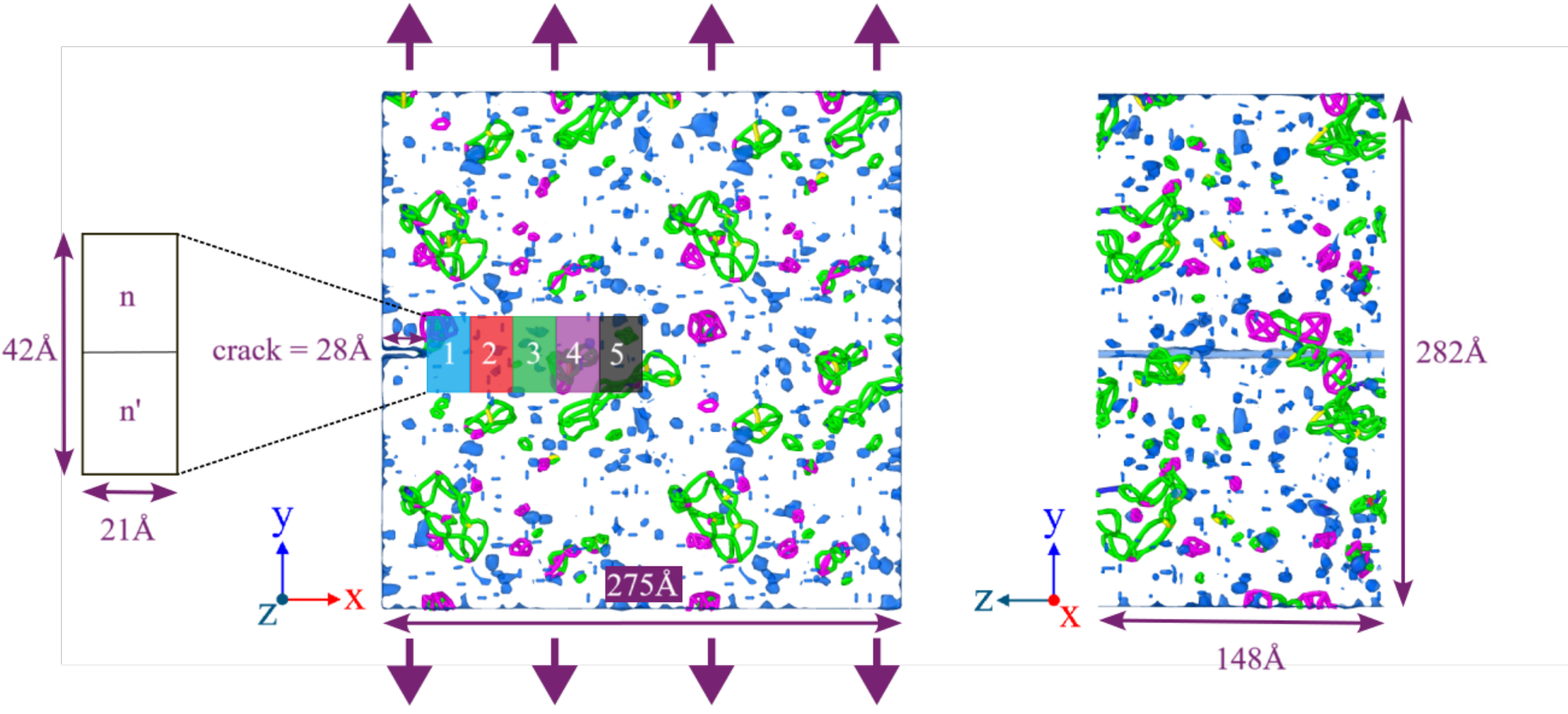}
    \caption{Three-dimensional MD model for simulating fracture under tensile loading with a pre-crack, showing the $x$--$y$ projection on the left, and $y$--$z$ projection on the right for (001) orientation; five colored boxes indicate subvolumes used to model the material's resistance to crack propagation.}
    \label{fig: Mode-i}
\end{figure}

After replicating each irradiated sample, a relaxation step was performed to ensure system stability under desired conditions. Initially, the atomic configuration was energy-minimized to reduce potential energy and relax internal stresses. This was followed by equilibration using a canonical NPT ensemble for 100 ps at zero pressure and a temperature of 300~K, allowing the system to adjust both atomic positions and box dimensions. This step allowed the system to equilibrate under the target conditions before proceeding with crack creation and the subsequent deformation process.

\begin{table}[t!]
\centering
\caption{Specification of the simulation cells used in crack-propagation simulations (after irradiation, replication and crack insertion), including Schmid factors for slip ($m^{\rm slip}_{\rm max}$) and twinning ($m^{\rm twin}_{\rm max}$).}

\vspace{2ex}

\renewcommand{\arraystretch}{1.25}
{\small
\begin{tabular}{ccccccc}
\hline
Orientation & Crack geometry & Loading axis & $m^{\rm slip}_{\rm max}$ & $m^{\rm twin}_{\rm max}$ & Dimensions \\[-1.25ex]
 & $\{\text{plane}\}\langle\text{front}\rangle$ & & & & (\AA$^3$) \\
\hline
(001) & $\{010\}\langle001\rangle$ & $[010]$ & 0.408 & 0.236 & $275 \times 282 \times 148$ \\
(011) & $\{01\bar{1}\}\langle011\rangle$ & $[01\bar{1}]$ & 0.408 & 0.471 & $275 \times 284 \times 147$ \\
(111) & $\{\bar{1}01\}\langle111\rangle$ & $[\bar{1}01]$ & 0.408 & 0.471 & $274 \times 284 \times 147$ \\
\hline
\label{Tab:crack_models_replicated}
\end{tabular}
}
\end{table}

In the crack propagation simulations, the periodic boundary conditions were applied along the thickness direction. In contrast, non-periodic boundaries were implemented along the $x$- and $y$-directions that correspond, respectively, to the applied loading and crack propagation directions \citep{Krull2011, ZENG2018120, XIE2023, Ustrzycka2025}.
The top and bottom atomic layers, each with a thickness of $2a$ (where $a \!=\! 3.526\,\text{\AA}$ is the lattice constant), were constrained to move as rigid bodies during the loading process, as a way of imposing tensile deformation of the sample. Additionally, constraints were enforced on the atoms located at the right and left edges of the sample (within layers of a thickness of $1a$), such that these atoms were restricted from moving in the $x$-direction but allowed free movement in the $y$- and $z$-directions. This boundary condition was selected based on a systematic evaluation of different boundary configurations presented in \citet[][Appendix~B]{Ustrzycka2025}, which shows that, due to the pronounced plasticity of the analysed alloy, imposing lateral constraints effectively minimizes residual deformations and prevents necking at the sample edge.

Following equilibration and the application of boundary conditions, the NVT ensemble was applied to ensure a stable environment for accurate simulation of material deformation at 300~K, and the primary crack, with a length of 28~\text{\AA}, approximately 1/10 of the total specimen length, was introduced by removing the bonds between atoms in the mid-left region of the sample (see Fig.~\ref{fig: Mode-i}). This crack-introduction method does not introduce artificial stress concentrations or structural artifacts before mechanical loading. The adopted crack-to-length ratio falls within the range typically reported in MD studies of tensile and fracture tests, where the initial crack length is chosen as a fraction of the overall specimen length (commonly 1/10, 1/5, 1/4, or 1/3) \citep{Ma2022}.

The velocity of the top and bottom rigid layers, as described above, is prescribed to achieve a global strain rate of $10^9$ s$^{-1}$, with the layers displaced with the same velocity of 0.142~\AA/ps in opposite directions along the $y$-axis to apply tensile deformation, while preserving symmetry in the simulation setup. This strain rate is commonly used in MD simulations of crack propagation \citep{WANG2015,lu2020cohesive, XIE2023}. A time step of 1 fs was employed, consistent with previous MD investigations \citep{Krull2011, wang2021molecular}.

In addition to the reference strain rate of $10^9$ s$^{-1}$, simulations were also carried out for the (001) and (011) orientations at lower rates of $10^8$ s$^{-1}$ and 5~$\times~10^7$ s$^{-1}$ to investigate the influence of the strain rate on main deformation mechanisms. Due to high computational cost, these cases were limited to a reduced overall strain range of up to 6\%. 

To ensure statistical robustness, crack-propagation simulations were performed on statistically independent irradiated configurations for each crystallographic orientation (four for (011) and three for (001) and (111)). These independent realizations were generated by rotating the samples or by cutting and reassembling different segments of the irradiated volume, thereby modifying the spatial arrangement of defects with respect to the crack front without altering the overall damage level. This procedure ensures that the crack does not accidentally initiate or propagate through an atypical defect cluster, and that the reported responses represent robust trends rather than artifacts of a particular defect configuration.Each realization was treated individually, and the final characteristics (along with the corresponding standard deviations) reported in Section~\ref{sec: Section4.3} are obtained by averaging the values obtained from individual simulations.

For future use, five distinct regions near the crack tip (indicated by the colored boxes in Fig. \ref{fig: Mode-i}) are defined, each in dimensions $21\times42\times148$ \text{\AA}\(^3\). These regions correspond to subvolumes where the material's resistance to crack propagation is examined. The details of this approach will be discussed in Section \ref{sec: Section4}.

%% file: sections/3_Physical_mechanisms.tex
\section{Mechanisms of crack growth across crystallographic orientations}
\label{sec:Section3}

This section presents a qualitative analysis of crack propagation in unirradiated and irradiated single crystals with the (001), (011), and (111) orientations specified in Table~\ref{Tab:crack_models_replicated}. In the case of the irradiated samples, the analysis is performed for the highest dose of 0.152 dpa (400 collision cascades).

Radiation-induced defect structures constitute a unique class of microstructural features, arising from highly energetic collision cascades and evolving under profoundly non-equilibrium thermodynamic conditions. Their characteristic morphologies and interactions with crystalline defects fundamentally alter the balance between plastic deformation and the material’s susceptibility to brittle fracture. Understanding the mechanisms by which such defect populations mediate crack initiation and advance is central to resolving the problem of radiation-induced hardening and embrittlement, as well as the accompanying transition from ductile to brittle behavior. 

The fundamental deformation mechanisms are common to all orientations; however, crystallographic symmetry and the spatial arrangement of radiation defects dictate their activation pathways, such that distinct processes emerge as dominant in different orientations. The following subsections discuss these mechanisms, considering also their contribution to the fracture behavior of irradiated materials.

\subsection{Pristine samples: orientation-dependent crack tip plasticity mechanisms}
\label{sec:Section3.1}
\vspace{0.2cm}
Plastic deformation near the crack tip in pristine samples is governed by activation of crystallographically favorable slip systems, determined by the orientation relative to the applied load. 
Recall that the Schmid factor for slip is identical for the three considered orientations (Table~\ref{Tab:crack_models_replicated}), hence the observed differences between the corresponding behaviors result from more subtle aspects, such as the orientation of the \{111\} planes with respect to the crack plane and crack front, which govern dislocation emission and interaction.

\begin{figure}[b!]
    \centering
    \includegraphics[scale=0.65]{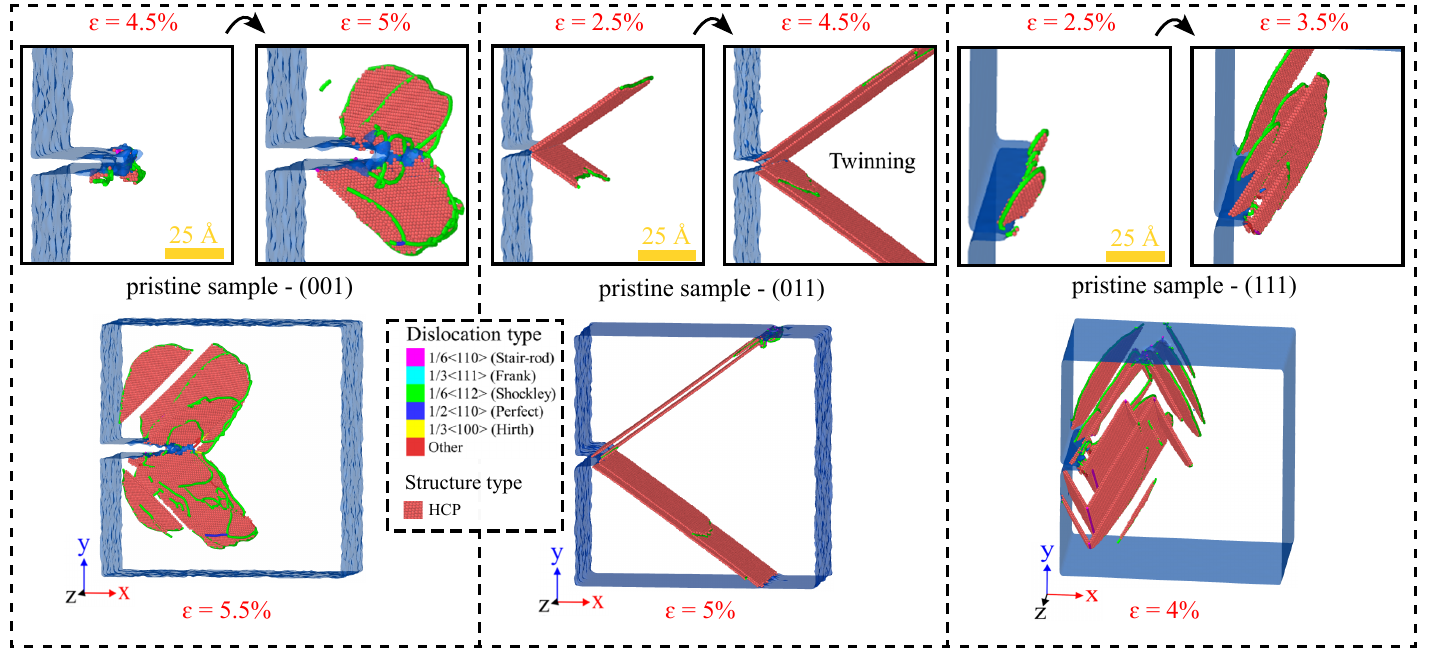}
 \caption{Orientation-dependent crack-tip deformation in pristine (001), (011), and (111) samples under mode-I loading. The top row shows close-ups of crack-tip regions, highlighting dislocation activity and twinning (at the early stage of deformation, twinning is activated only for the (011) orientation). The bottom row presents full-sample views at later stages.}
    \label{fig:purecompare}
\end{figure}

Fig.~\ref{fig:purecompare} presents atomistic snapshots under mode-I tensile loading, capturing dislocation emission, extension of stacking-faulted regions via Shockley partial glide (stacking fault emission), and twinning.

In (001), nucleation of Shockley partials is delayed (note that the first (001) snapshot in Fig.~\ref{fig:purecompare} corresponds to $\varepsilon$=4.5$\%$ which is higher than the first snapshots shown for the other orientations), and the inelastic deformation region is initially confined to the very vicinity of the crack front. 
In the (011) orientation, the initially emitted stacking faults develop into two symmetric twins that nucleate at the crack front through a mechanism discussed by \citet{Warner2007} and \citet{ANDRIC2018144}. This behavior is promoted by the geometry of the \{111\} planes intersecting along the crack front aligned with the [011] direction, enabling efficient activation of the twin‑related partial dislocation network. Importantly, twinning is not unique to the (011) orientation; stacking faults and partial dislocation activity occur in all orientations at later stages, although they are significantly less pronounced in the (001) and (111) cases. The apparent predominance of twin‑like features in (011) reflects orientation-dependent differences in slip system activation, resolved shear stress, and activation barriers.

In the (111) orientation, multiple glide planes are readily activated, leading to extensive partial dislocation activity, while twin formation from the crack front is limited under the present loading conditions. These differences in the dominant deformation mechanism (dislocation glide versus twinning), the onset of nucleation events, and the spatial extent of stacking-faulted regions around the crack tip define the orientation-specific development of the plastic zone.

\subsection{Irradiated sample: (001) orientation}
\label{sec:Section3.2}
\vspace{0.2cm}
\begin{figure}[t!]
    \centering
    \includegraphics[scale=0.325]{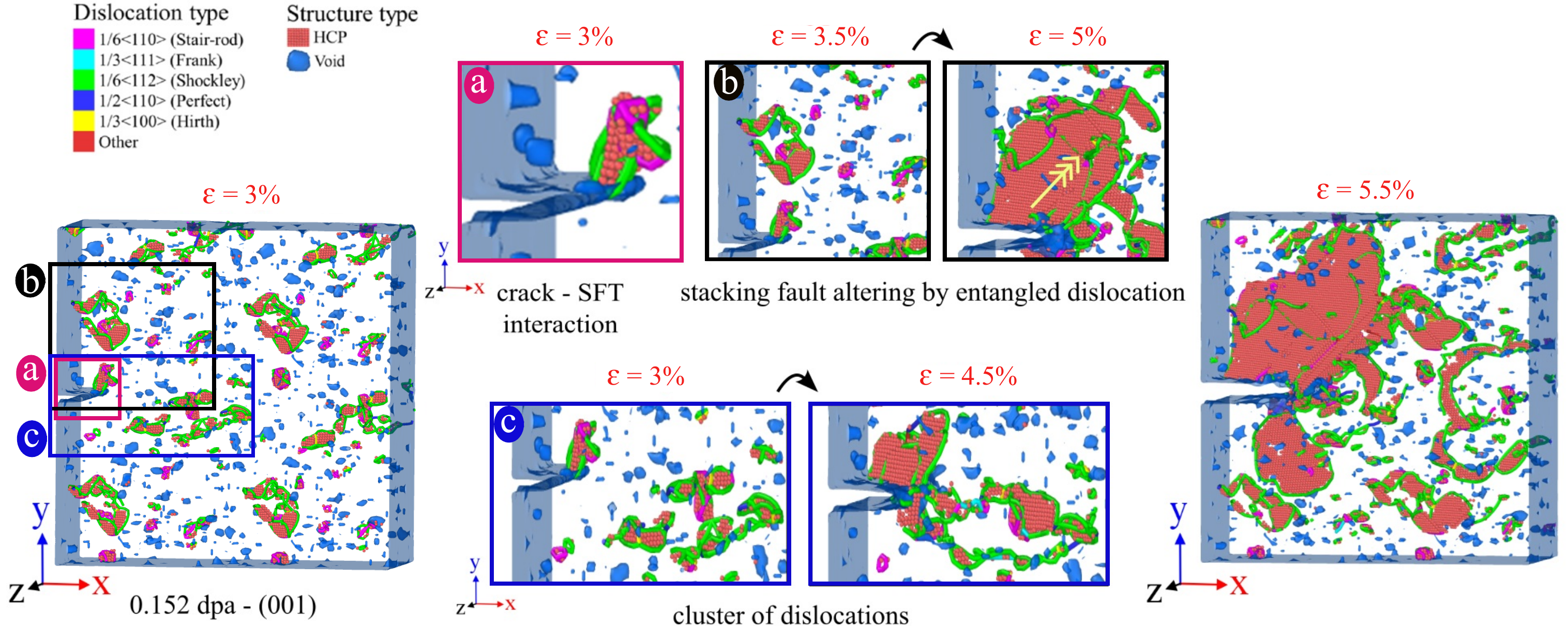}
\caption{Initial crack evolution in a (001)-oriented sample irradiated to 0.152~dpa. Side images show crack positions at $\varepsilon$=3$\%$  and $\varepsilon$=5.5$\%$, while middle frames highlight active mechanisms near the crack tip: (a) interaction of crack tip with radiation-induced SFT ($\varepsilon$=3$\%$; purple frame), (b) absorption of entangled dislocations into the crack-tip plastic zone ($\varepsilon$ from 3.5\% to 5\%; black frames), and (c) interaction with dislocation clusters that impede stacking fault emission (i.e., the extension of stacking-faulted regions via glide of Shockley partials) ($\varepsilon$ from 3\% to 4.5\%; blue frames). Yellow arrow indicates the direction of stacking fault emission.}
    \label{fig:001dxacna}
\end{figure}
\begin{figure}[t!]
    \centering
    \includegraphics[scale=0.135]{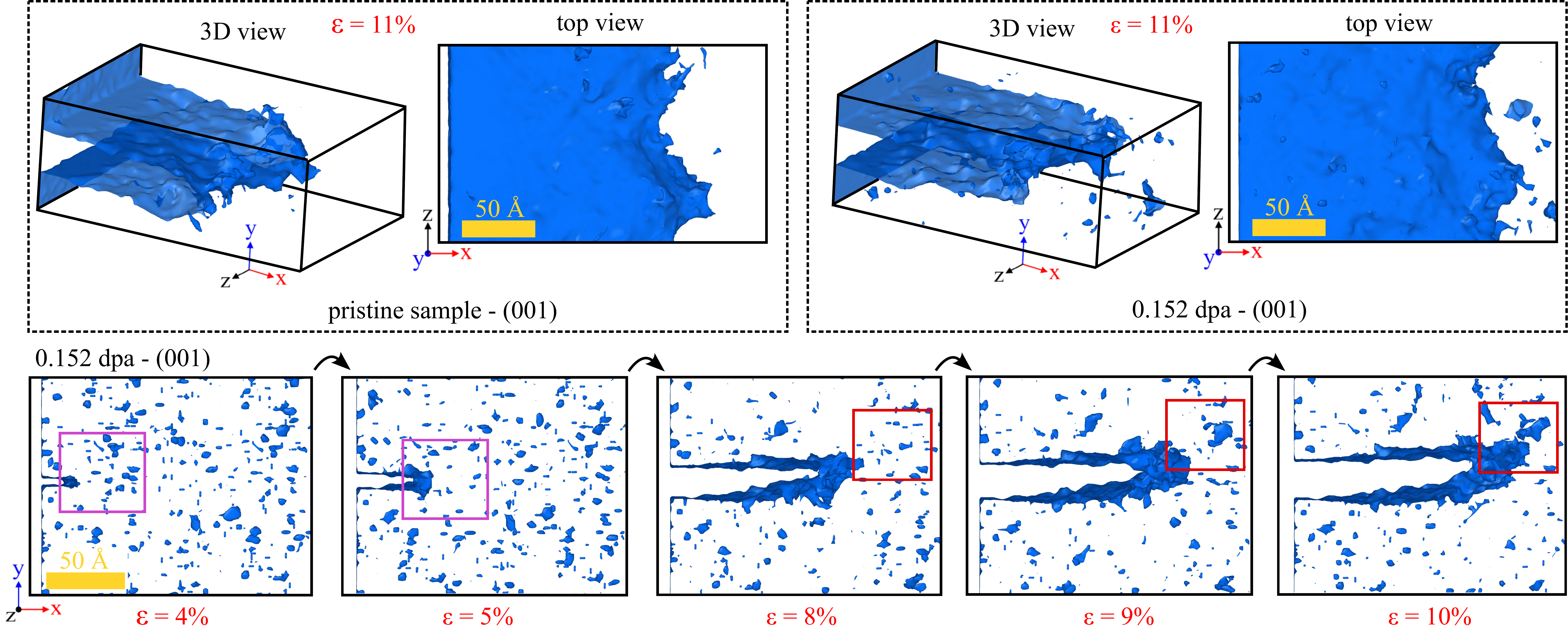}
\caption{Crack propagation in a (001)-oriented sample irradiated to 0.152~dpa, showing the role of radiation-induced voids (dislocations and stacking faults are not shown). Top row: comparison with the pristine sample at the overall strain $\varepsilon$=11$\%$ reveals similar crack lengths, with local deviations of crack shape linked to void--crack interactions in the irradiated sample. Bottom row: sequential snapshots ($\varepsilon$ varying from  4\% to 10\%) highlight void absorption at the crack tip (purple rectangles) and subsequent void coalescence ahead of the crack (red rectangles).}
    \label{fig:001compare}
\end{figure}

Fig. \ref{fig:001dxacna} shows the early evolution of a crack in a (001)-oriented sample irradiated to 0.152~dpa, highlighting key interactions with radiation-induced defects. Note that the observed mechanisms are consistent across all analysed doses (0.008, 0.038, 0.076, and 0.152~dpa). Fundamental mechanisms of a single defect–crack interaction governing crack propagation were previously shown by \citet{Ustrzycka2025} for the (001) orientation.

In Fig.~\ref{fig:001dxacna}, frame (a) (pink box) shows the crack front interacting with radiation-induced SFTs, which act as barriers to stacking-fault emission. Frames (b) (black box) highlight the absorption of radiation-induced entangled dislocations into the crack-tip plastic zone, altering slip symmetry. Frame (c) (blue box) shows the interaction with dislocation clusters that impede the propagation of stacking faults, collectively producing an asymmetric plastic field around the crack tip.

Radiation-induced voids further influence crack morphology (Fig.~\ref{fig:001compare}). The blue regions in the visualization correspond to isosurfaces that capture both crack cavities and radiation-induced voids, with dislocation lines omitted to emphasize crack morphology and its interaction with voids. At low strains ($\varepsilon$ from 4\% to 5\%, purple frames), pre-existing voids locally reduce material strength and are absorbed by the advancing crack, accelerating its growth. At higher strains ($\varepsilon$ from 8\% to 10\%, red frames), void coalescence ahead of the crack connects weakened regions, forcing the crack along a more tortuous path. These results confirm and complement our previous findings \citep{Ustrzycka2025}.
\subsection{Irradiated sample: (011) orientation}
\label{sec:Section3.3}
\vspace{0.2cm}
\begin{figure}[b!]
    \centering
    \includegraphics[scale=0.33]{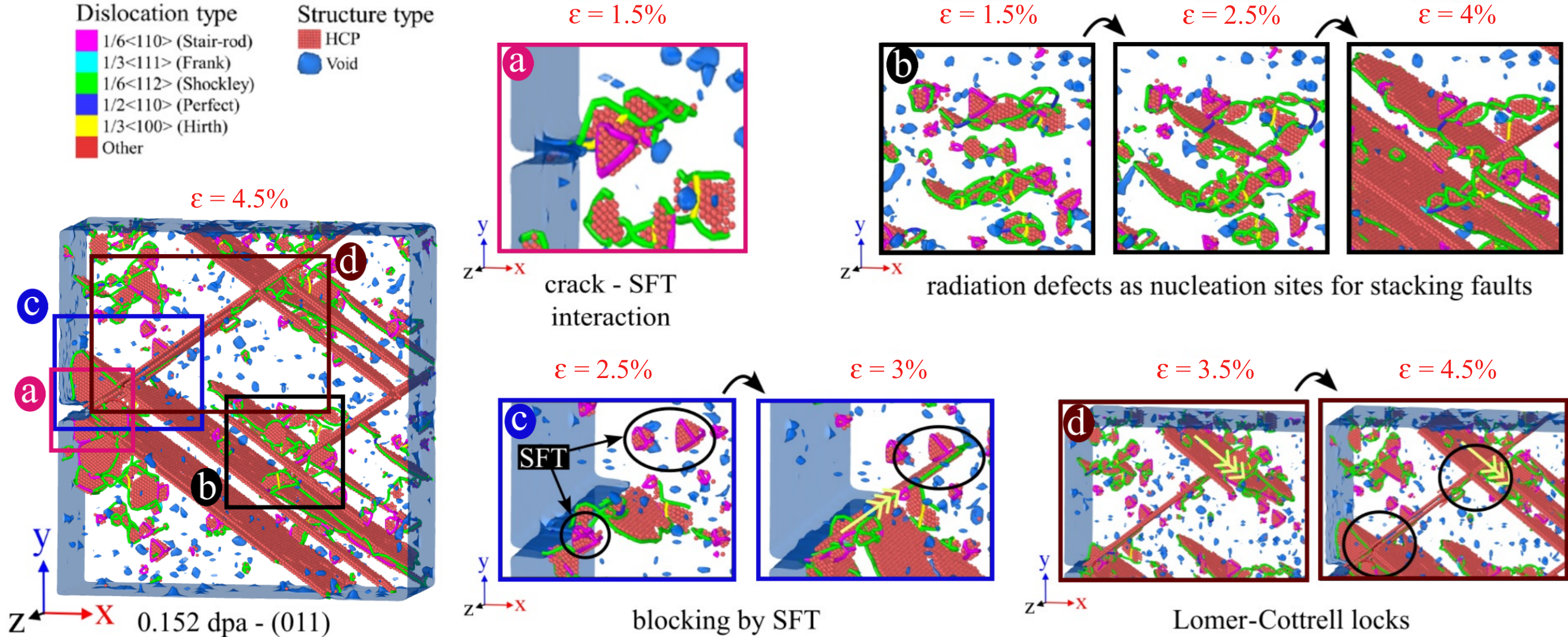}
    \caption{Crack evolution in a (011)-orientated sample irradiated to 0.152~dpa. Zoomed-in snapshots show key deformation mechanisms: (a) initial interaction between the crack front and a radiation-induced SFT ($\varepsilon$=1.5$\%$; purple frame), (b) nucleation of plasticity at radiation-induced defects ($\varepsilon$=1.5--4$\%$; black frames), (c) obstruction of stacking fault evolution by SFTs ($\varepsilon$=2.5--3$\%$; blue frames), and (d) formation of Lomer--Cottrell locks at stacking fault intersections near the crack tip ($\varepsilon$=3.5--4.5$\%$; brown frames). Yellow arrows indicate the direction of stacking fault emission.}
\label{fig:011dxacna}
\end{figure}

\begin{figure}[t!]
    \centering
    \includegraphics[scale=0.13]{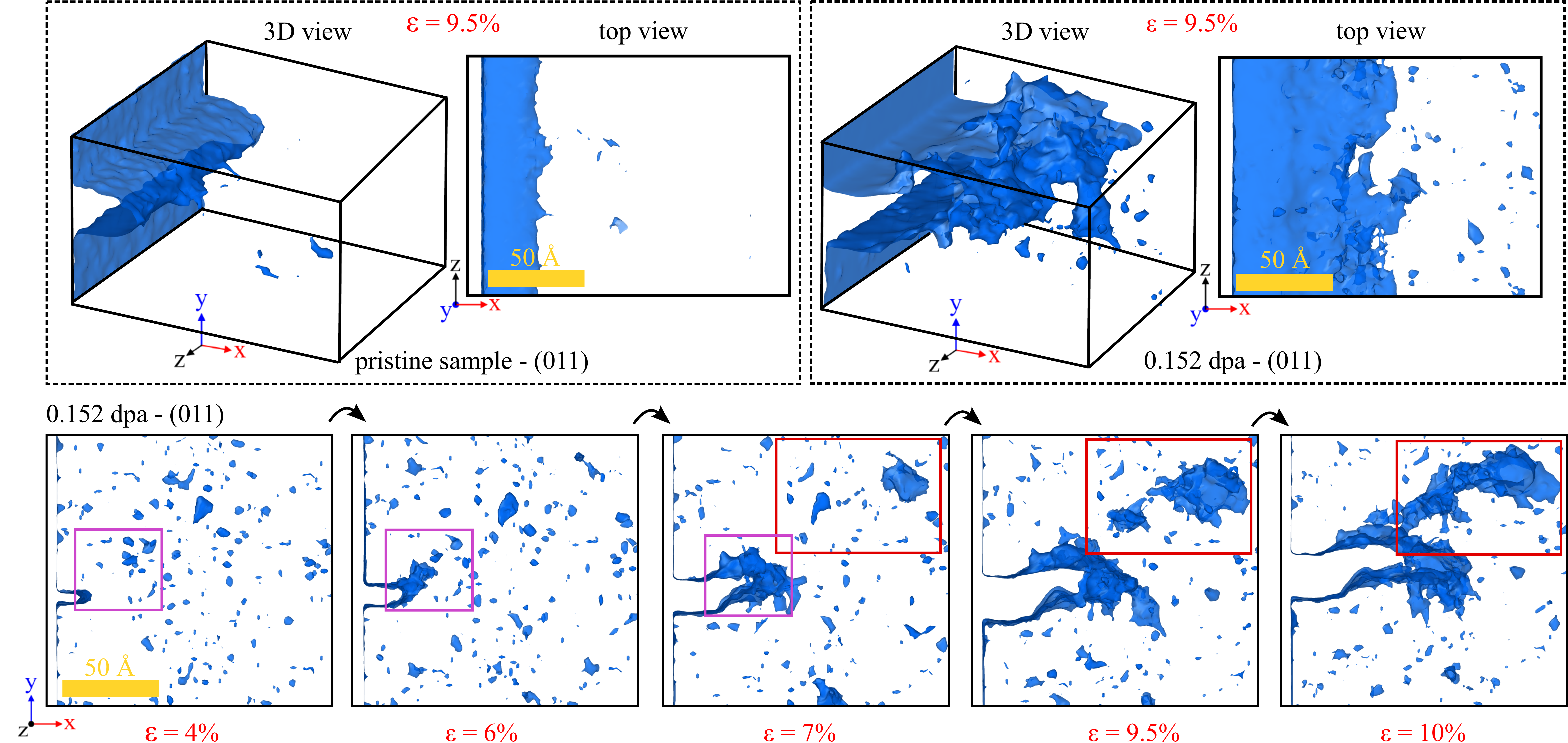}
    \caption{Role of radiation-induced voids on crack propagation in (011)-oriented sample irradiated to 0.152~dpa (dislocations and stacking faults are not shown). 
Top row: comparison of crack length and morphology between pristine and irradiated samples at the same strain level ($\varepsilon$=9.5$\%$) highlights void-driven crack advance in the irradiated case. 
Bottom row: sequence from $\varepsilon$=4$\%$ to 10$\%$ shows initial absorption of voids near the crack tip (purple rectangles) up to $\varepsilon$=7$\%$, followed by void accumulation and coalescence that disrupt the crack front and deflect its path (red rectangles).}
    \label{fig:011compare}
\end{figure}

In a (011)-oriented sample irradiated to 0.152~dpa, shown in Fig.~\ref{fig:011dxacna}, crack-tip plasticity develops through successive interactions with radiation-induced defects, fragmenting the coordinated twinning observed in the pristine material. In the unirradiated state, as illustrated in Fig.~\ref{fig:purecompare}, 
the geometrical arrangement of the crack front within two \{111\} planes promotes extended stacking faults and twins, leading to effective crack-tip blunting. Irradiation disrupts this mechanism.

At early times ($\varepsilon$=1.5$\%$; purple frame (a) in Fig.~\ref{fig:011dxacna}), the crack front interacts with SFTs, triggering localized slip but altering the conditions for fault emission. Between $\varepsilon$=1.5$\%$ and 4$\%$ (black frames (b)), deformation initiates from dispersed radiation defect sites, producing a fragmented plastic zone.

At $\varepsilon$=2.5--3$\%$ (blue frames (c)), SFT blocks dislocation emission and limits the extension of stacking-faulted regions, suppressing forward plasticity along primary slip systems. Later ($\varepsilon$=3.5--4.5$\%$; brown frames (d)), intersecting faults form Lomer--Cottrell locks that restrict slip redistribution. Overall, plastic flow becomes spatially discontinuous and asymmetric.

Similar to the (001) orientation, radiation-induced voids in the (011) orientation promote crack advance by two successive mechanisms: at low strains ($\varepsilon$ from 4\% to 10\%) they are absorbed by the crack front, while at higher strains ($\varepsilon$ from 4\% to 10\%) they coalesce ahead of it, forming weak links that deflect and accelerate propagation. The top row of Fig.~\ref{fig:011compare} shows that at $\varepsilon$=9.5$\%$ the irradiated (011) sample develops a longer, more tortuous crack path than in the unirradiated state, in contrast to the (001) orientation (see Fig.~\ref{fig:001compare}), where the crack length remains nearly unchanged by irradiation. 

Compared to (001), where the limited slip activity remains essentially unaffected by irradiation and allows relatively unhindered crack advance, the irradiated (011) sample shows suppression of twinning and reduced slip activity. Here, the defect-induced reduction of plasticity, combined with void-assisted crack growth, indicates a radiation-induced DBT, as confirmed by the analysis in Section~\ref{sec: Section4}.

The radiation-induced change in the deformation mechanisms observed in the (011) orientation appears to be primarily associated with interactions of partial dislocations and twins with the radiation-induced defects, including the formation of Lomer--Cottrell locks that restrict dislocation motion. At the same time, crystallographic orientation plays an important role by determining which plastic mechanisms, such as glide of dislocations or twinning, can be activated under stress. 
Together, these observations suggest that radiation-induced embrittlement reflects both the nature of defect–mechanism interactions and the accessibility of underlying plasticity modes, which vary with orientation. Consequently, similar embrittlement effects could emerge in other orientations if the local plastic mechanisms are similarly engaged. In particular, those orientations for which twinning is active are expected to be more prone to DBT.

\subsection{Irradiated sample: (111) orientation}
\label{sec:Section3.4}
\begin{figure}[b!]
    \centering
    \includegraphics[scale=0.28]{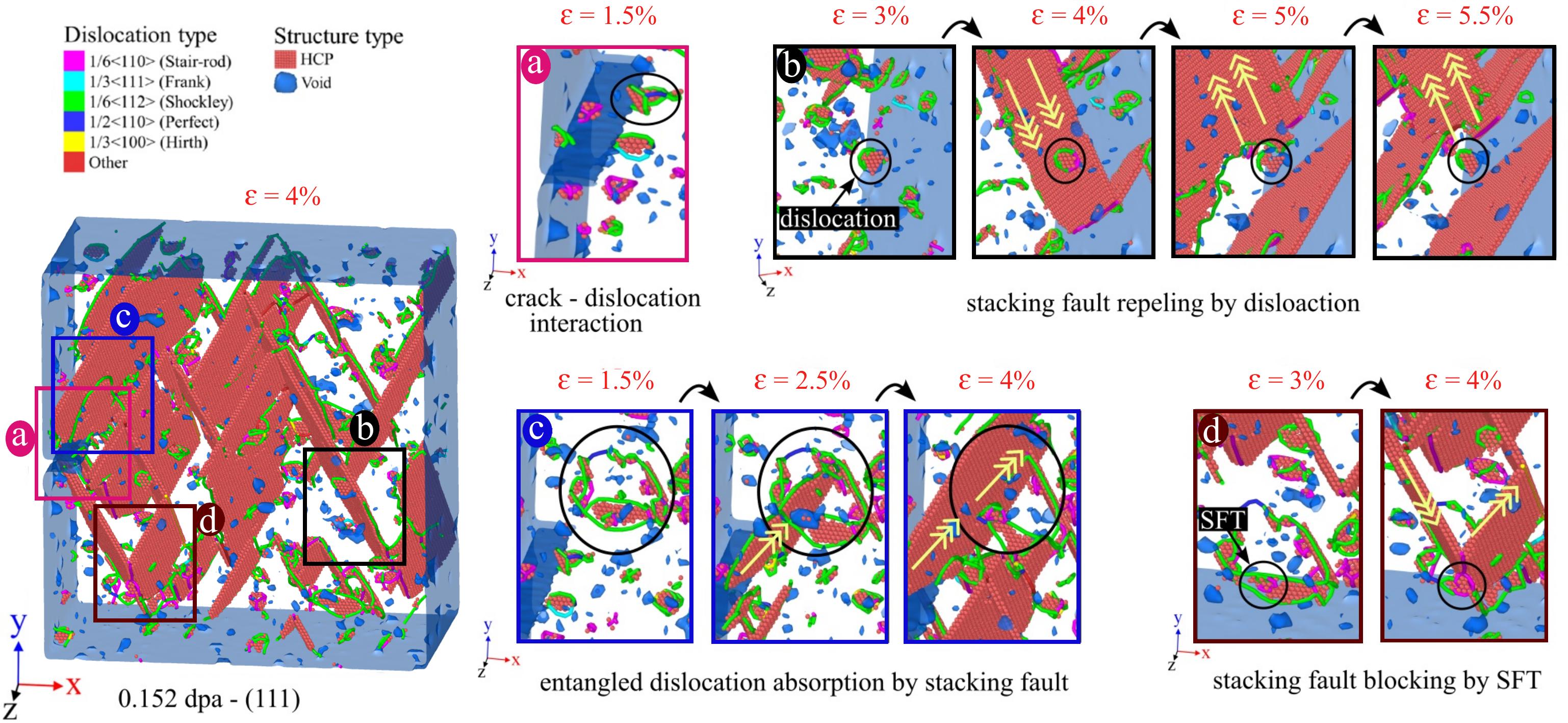}
\caption{Crack evolution in a (111)-oriented sample irradiated to 0.152~dpa, highlighting the interaction between stacking faults and radiation-induced defects. 
Zoomed-in snapshots illustrate key deformation mechanisms: (a)~initial interaction between the crack front and a radiation-induced dislocation loop ($\varepsilon$=1.5$\%$), (b)~deflection of stacking faults by radiation dislocations ($\varepsilon$=3--5.5$\%$), (c) absorption of entangled dislocations into stacking faults ($\varepsilon$=1.5--4$\%$), and (d) blocking of stacking fault extension by radiation SFT ($\varepsilon$=3--4$\%$). Yellow arrows indicate the direction of stacking fault emission.}
    \label{fig:111dxacna}
\end{figure}
\begin{figure}[t!]
    \centering
    \includegraphics[scale=0.12]{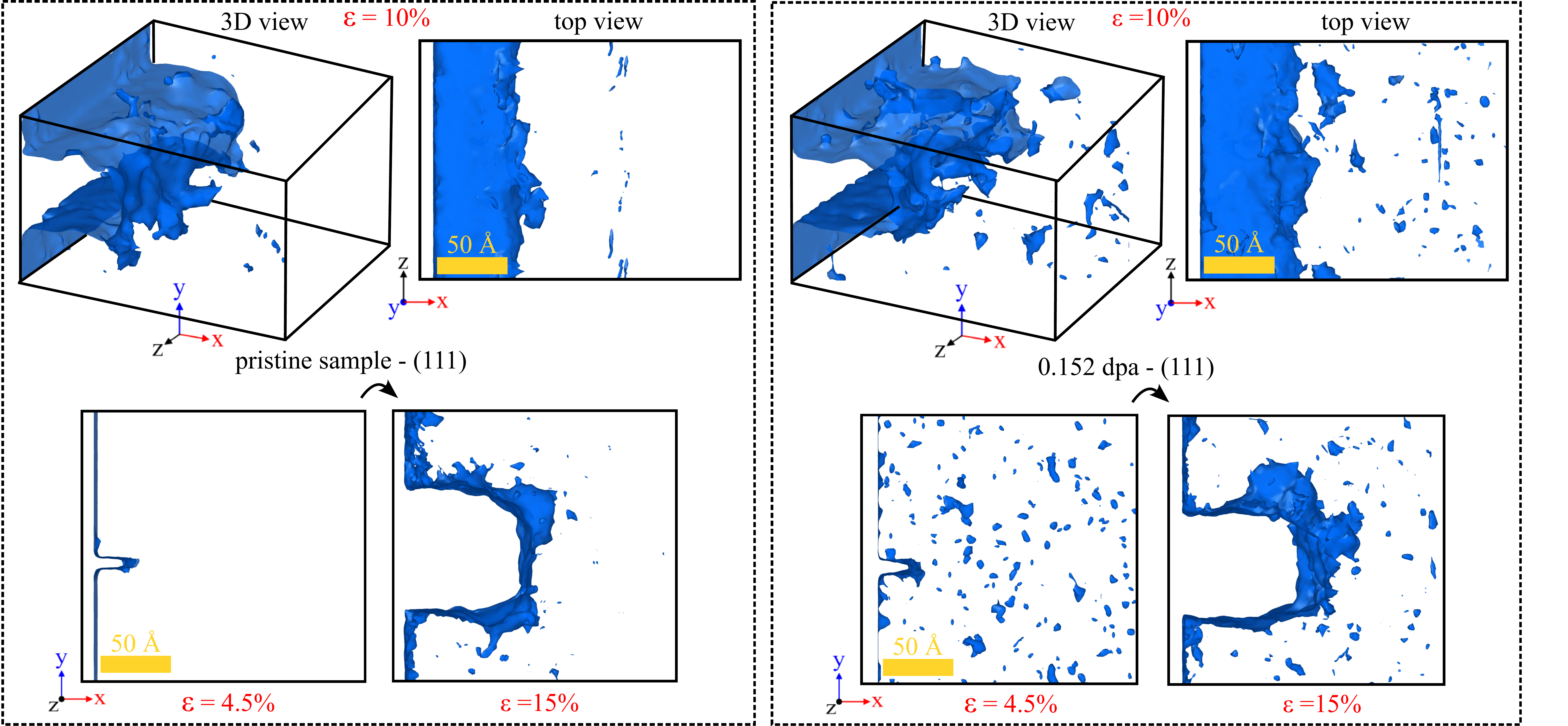}
    \caption{Comparison of crack behavior in a (111) orientation before and after irradiation (0.15~dpa), highlighting the limited influence of radiation-induced voids and the role of crack-tip blunting (dislocations and stacking faults are not shown). 
Top row: crack surface morphology in pristine (left) and irradiated (right) samples at the same strain level ($\varepsilon$=10$\%$) reveals enhanced plastic deformation and blunting. 
Bottom row: evolution of crack tip structure from $\varepsilon$=4.5$\%$ to 15$\%$ demonstrates sustained plasticity and delayed crack advance with crack tip blunting in both conditions.}
    \label{fig:111compare}
\end{figure}
\vspace{0.2cm}
In a (111)-oriented sample irradiated to 0.152~dpa, crack evolution is dominated by stacking fault emission and dislocation glide along multiple intersecting slip systems. The favorable crystallography ensures highly efficient and spatially distributed plasticity, which remains largely unaffected by radiation-induced defects, consistent with the pristine sample response (compare Fig.~\ref{fig:purecompare}).

Fig.~\ref{fig:111dxacna} illustrates these mechanisms. The full-field image (left) shows the irradiated sample at $\varepsilon$=4$\%$, while panels (a)--(d) highlight key defect--fault interactions. At early crack advance ($\varepsilon$=1.5$\%$, purple box), the front interacts with dislocation loops, initiating stacking fault emission. Subsequently ($\varepsilon$=3--5.5$\%$, black box), emitted faults encounter loops that locally repel propagation but do not impede plastic flow. Radiation-entangled dislocations are then absorbed into stacking faults ($\varepsilon$=1.5--4$\%$, blue box), integrating defects into the plastic front. Finally, limited fault blockage by SFTs occurs ($\varepsilon$=3--4$\%$, brown box), yet continuous deformation is maintained.

Fig.~\ref{fig:111compare} shows the limited influence of radiation-induced voids in the (111) orientation. At the strain of $\varepsilon$=10$\%$, the crack morphology in the irradiated sample is similar to that of the pristine case, with both exhibiting pronounced blunting, indicating that voids have little effect on the local crack-tip response. Snapshots at $\varepsilon$=4.5$\%$ and $15\%$ (lower row) confirm the minimal impact of voids on crack propagation in this orientation.

\subsection{Dislocation density evolution during crack propagation}

\begin{figure}[t!]
    \centering
    \includegraphics[width=1.02\textwidth]{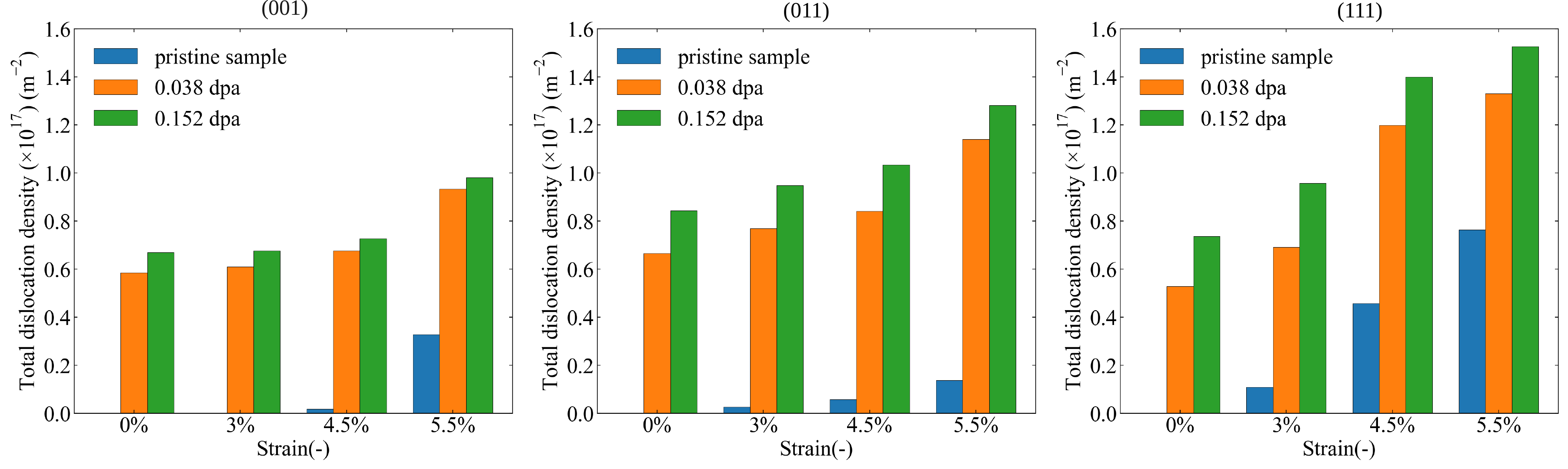}
    \caption{
Evolution of total dislocation density during crack propagation as a function of applied strain for pristine and irradiated samples in the three crystallographic orientations.}
\label{fig:dislocation_comparing}
\end{figure}

Fig.~\ref{fig:dislocation_comparing} presents the evolution of the total dislocation density during crack propagation for the three analyzed crystallographic orientations, evaluated in both pristine and irradiated states (0.038 and 0.152 dpa). Across all orientations, at zero strain, irradiation establishes the initial dislocation content, reflecting the presence of radiation-induced defects. As deformation progresses and the crack advances, all orientations show a systematic increase in dislocation density, reflecting continued crack-tip emission and interactions with the pre-existing defect population.

The dislocation density evolution exhibits a coupled dependence on the crack geometry (orientation) and irradiation dose.

The (001) configuration develops the lowest dislocation densities across the examined strain range. This behavior is consistent with the restricted activity of operative slip systems available to accommodate crack-tip plasticity, as discussed in Section~\ref{sec:Section3.2}.  Even at $\varepsilon$=5.5$\%$, the dislocation density remains below $\sim 1.0 \times 10^{17}\mathrm{m^{-2}}$, underscoring the strongly constrained plastic response characteristic of this orientation.

The (011) orientation displays an intermediate increase in dislocation density under irradiation. In the pristine state, dislocations nucleated at the crack tip move away efficiently along the available glide systems, leaving the near-tip region with only a small residual dislocation content. Irradiation fundamentally changes this mechanism. Radiation-induced defect clusters act as strong obstacles, reducing dislocation mean free paths, promoting trapping, and leading to localized dislocation accumulation, consistent with established observations of radiation-induced embrittlement. As detailed in Section~\ref{sec:Section3.3}, this leads to a substantial elevation of local dislocation density for both irradiation levels.

The (111) orientation shows the highest dislocation accumulation, reaching $\sim 1.5 \times 10^{17},\mathrm{m^{-2}}$ at 5.5$\%$ strain for 0.152 dpa. This pronounced buildup is consistent with the high symmetry and multiplicity of slip systems intersecting the ⟨111⟩ crack-front direction. These crystallographic conditions favor extensive crack-tip plasticity, dislocation emission, and enhanced cross-slip, as discussed in Section~\ref{sec:Section3.4}. Irradiation amplifies these trends by increasing dislocation--defect interactions and limiting the escape of emitted dislocations, resulting in the highest defect-assisted storage among all examined orientations.

The results indicate that irradiation amplifies dislocation storage in a strongly orientation-dependent manner. These results correlate with the distinct crack-tip deformation mechanisms.

\subsection{Assessing strain-rate effects on radiation-induced fracture mechanisms}
\vspace{0.2cm}
\begin{figure}[b!]
    \centering
    \includegraphics[scale=0.152]{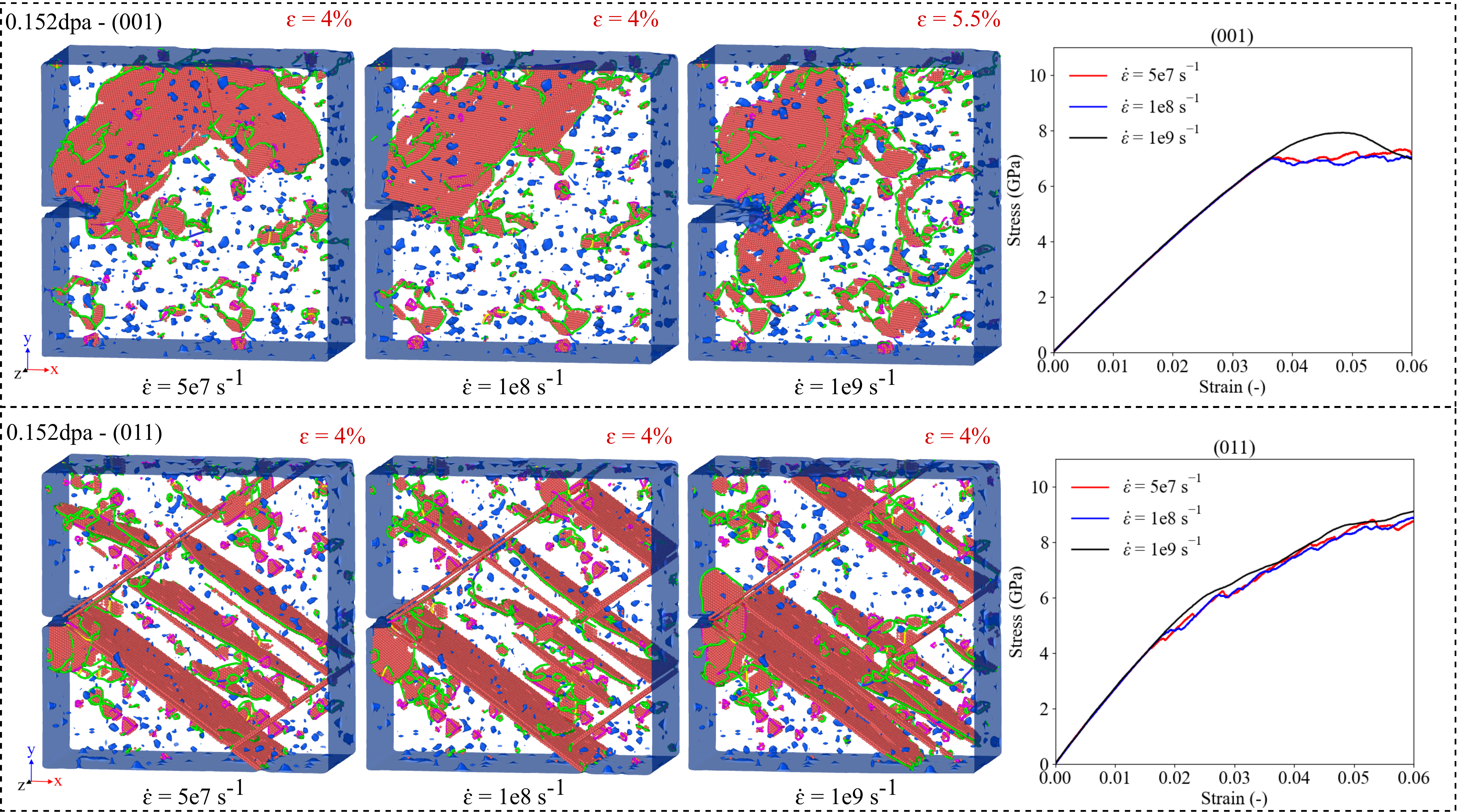}
    \caption{Strain rate-dependent crack propagation in irradiated samples for two orientations: (001) (top row) and (011) (bottom row). Simulations are conducted at ($\dot{\varepsilon}$=$5 \times 10^7$~$\text{s}^{-1}$), \( 10^8~\text{s}^{-1} \), and \( 10^9~\text{s}^{-1} \). Morphologies are captured at ($\varepsilon$=4$\%$) except for the (001) sample at \(10^9~\text{s}^{-1}\), where ($\varepsilon$=5.5$\%$) was needed due to delayed plasticity. The consistent appearance of dislocation clusters, void coalescence, and defect-stacking fault interactions confirms the convergence of underlying physical mechanisms across strain rates.}
    \label{fig:strainrate}
\end{figure}
The fracture mechanisms identified in Sections~\ref{sec:Section3.2}, \ref{sec:Section3.3}, and~\ref{sec:Section3.4} indicate that radiation-induced defects markedly influence crack-tip plasticity via orientation-dependent interactions with evolving deformation fields. These mechanisms were characterized under the high strain-rate conditions typical of MD simulations ($\dot{\varepsilon}$=$10^9$~\text{s}$^{-1}$).
However, the physics of dislocation nucleation, defect mobility, and crack-tip shielding is known to be sensitive to the applied deformation rate \citep{zhang2017molecular, ZHAO2025111667}. At elevated rates, reduced time for defect relaxation may suppress thermally activated processes, alter slip activity, and stabilize radiation-induced obstacles that would otherwise be absorbed or bypassed.

It is therefore essential to verify whether the main mechanisms remain operative across a broader range of loading rates. Demonstrating their rate-independence is crucial for confirming their physical relevance beyond the constraints of MD-specific conditions.

The simulation set at $\dot{\varepsilon}$ = 10$^9$~\text{s}$^{-1}$ was extended to include two additional loading rates: \( 10^8~\text{s}^{-1} \) and \( 5 \times 10^7~\text{s}^{-1} \), for the (001) and (011) orientations, as shown in Fig.~\ref{fig:strainrate}. The (111) orientation was not considered, as dominant plasticity in this case renders the effect of radiation-induced defects negligible. Due to high computational cost, the simulations were performed over a reduced strain range up to $\varepsilon$ = 6$\%$. Crack-tip morphologies were evaluated at $\varepsilon$=4$\% $, with the exception of the (001) sample at $\dot{\varepsilon}$=10$^9$~\text{s}$^{-1}$, where delayed plasticity required extension to $\varepsilon$=5.5$\%$. 

Note that the strain rates accessible in MD simulations ($\sim10^{8}$–$10^{9}$ s$^{-1}$) remain many orders of magnitude higher than those relevant in experiments (<10$^{3}$ s$^{-1}$). At such high rates, the time available for defect relaxation, dislocation escape from irradiation-induced obstacles, and other rate-dependent processes is strongly limited, leading to elevated apparent flow stresses and delayed dislocation nucleation \citep{Fan2021}.

Examination of Fig.~\ref{fig:strainrate} shows that variations in strain rate primarily influence the onset and kinetics of crack-tip plasticity, while the overall crack-propagation morphology remains comparable across the investigated conditions.

For the (001) orientation, the highest applied strain rate ($10^{9}$ s$^{-1}$) shifts the onset of crack-tip plasticity to higher imposed strains and stresses compared to lower strain rates, indicating that rapid deformation delays dislocation nucleation at the crack tip. As the strain rate is reduced, plastic slip is activated at lower imposed strains, but the dominant crack-tip deformation mode remains unchanged. For the (011) orientation, which exhibits extensive slip activity, both the crack-tip deformation patterns and the stress–strain response show only weak sensitivity to strain rate within the investigated range. The results indicate that strain rate primarily controls the timing and extent of dislocation activity, rather than introducing qualitatively different crack-tip deformation mechanisms.

When extending the interpretation toward experimental strain-rate conditions,  the present results imply a reduction in the stresses and strains required to activate crack-tip plasticity, accompanied by increased dislocation mobility and a greater probability of dislocation annihilation and rearrangement. Such kinetic effects are expected to alter the density, continuity, and spatial extent of dislocation structures and voids, rather than introduce new mechanisms. As a consequence, local stress concentrations at the crack tip may be partially relaxed at lower strain rates. Importantly, the simulations indicate that radiation-induced defects remain the primary factors governing crack-tip plasticity by acting as persistent obstacles to dislocation motion and by modifying the local deformation fields. While the relative contributions of competing deformation processes are expected to change quantitatively under experimental loading rates, the defect-controlled nature of the crack-tip response identified here is therefore expected to remain relevant.

%% file: sections/4_Mechanical_responses.tex
\section{Quantitative assessment of fracture behavior}
\label{sec: Section4}
Building upon the qualitative analysis of deformation mechanisms presented in the previous section, this part offers a quantitative assessment of how irradiation affects fracture behavior in samples with (001), (011), and (111) orientations under mode-I tensile loading with a pre-existing crack. The evaluation is based on stress--strain responses, crack propagation characteristics, and T--S curves extracted from MD simulations.

\begin{figure}[b!]
    \centering
    \includegraphics[scale=0.3]{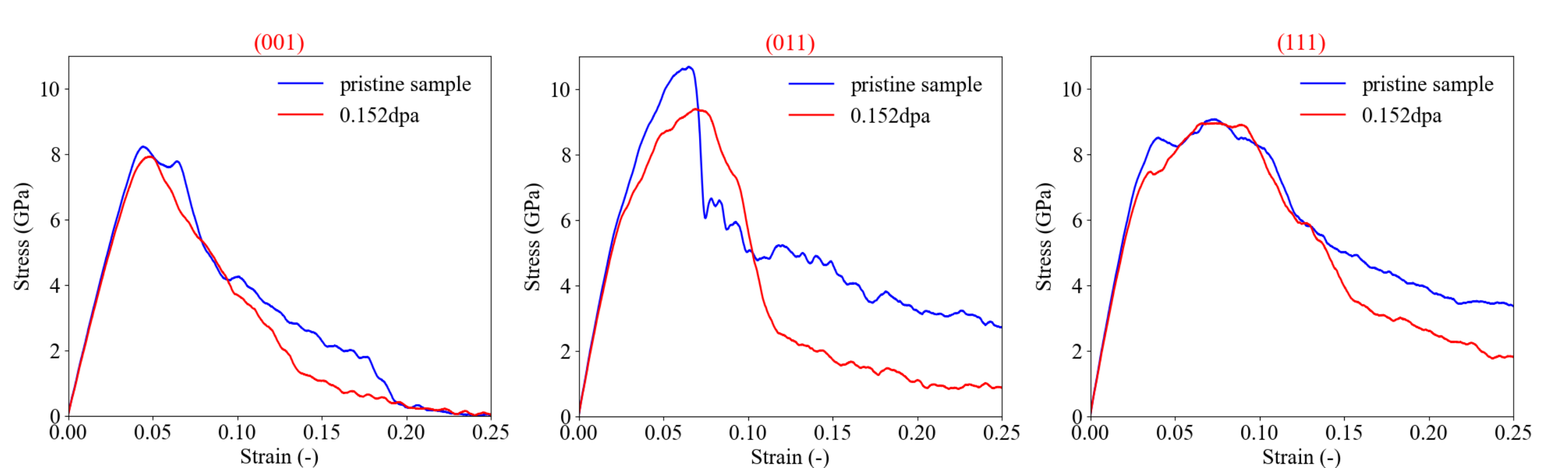}
    \caption{Orientation-dependent stress--strain responses for pristine and samples irradiated to 0.152~dpa.}
    \label{fig:SSPure}
\end{figure}
\subsection{Stress--strain response and crack length evolution}
\label{sec: Section4.1}
\vspace{0.2cm}
Fig.~\ref{fig:SSPure} presents orientation-resolved stress--strain curves for pristine and irradiated (0.152~dpa) samples. The extent of radiation-induced mechanical degradation is orientation dependent. Among the three orientations, the (011) orientation shows the most pronounced loss of ductility and a clear reduction in peak stress, indicating a radiation-induced DBT. As detailed in Section \ref{sec:Section3}, this transition arises from the combined effect of dislocation pinning by SFTs, the formation of Lomer--Cottrell locks, and the fragmentation of stacking faults, which together suppress plastic strain redistribution (Fig.~\ref{fig:011dxacna}). Additionally, void absorption and coalescence further facilitate crack propagation. The (111) orientation exhibits minimal sensitivity to irradiation: both yielding and early-stage hardening remain essentially unchanged, as multiple slip systems continue to accommodate plastic flow. As shown in Section \ref{sec:Section3}, absorption of defects into stacking faults and repulsive interactions between mobile and radiation-induced dislocations prevent strain localization, preserving ductility (Fig.~\ref{fig:111dxacna}). For the (001) orientation, where slip activity is intrinsically limited, irradiation produces only marginal additional effects, and the stress--strain response remains essentially unchanged (Fig.~\ref{fig:SSPure}). A detailed comparison of the crack-tip stress fields for all orientations and selected irradiation levels is provided in Appendix~\ref{sec:appendixA}, offering additional insight into the orientation-dependent plastic zone development and stress localization discussed here.

The serrated features in the curves arise from intermittent plastic events at the crack tip: the stress rises as crack propagation is temporarily blocked by entangled dislocations and stacking faults, followed by drops corresponding to barrier breaking and crack advance. Peaks also reflect successive stages of void nucleation and growth near the crack front, periodically relieving local stresses.
\begin{figure}[t!]
    \centering
    \includegraphics[scale=0.3]{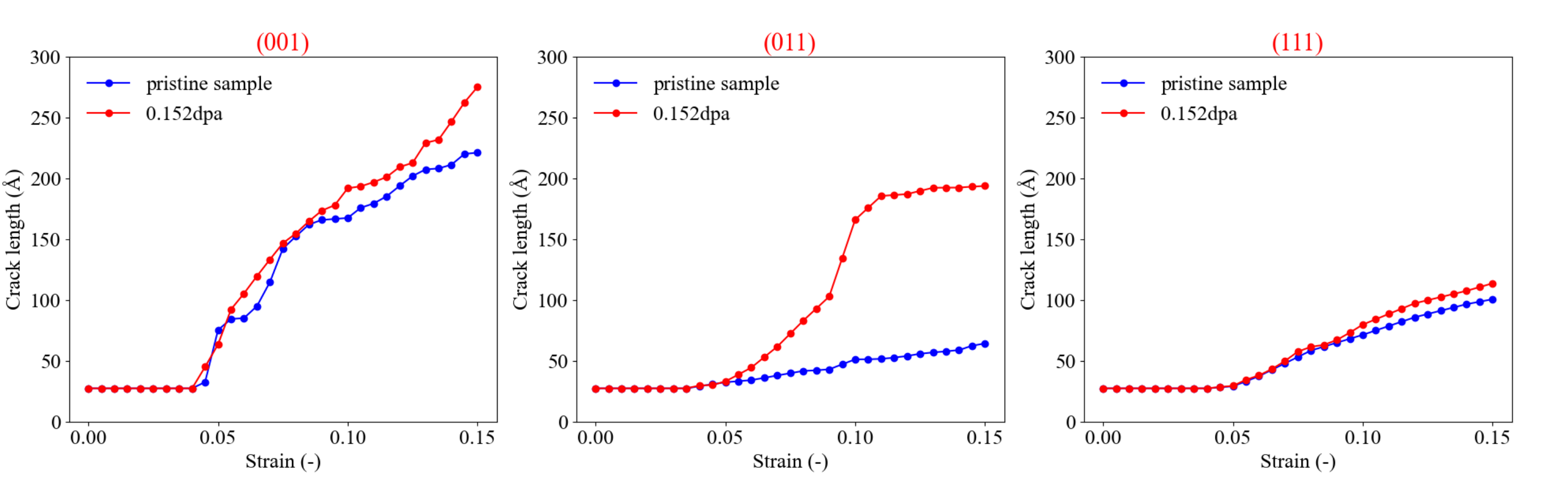}
    \caption{Crack length evolution as a function of strain for the (001), (011), and (111) orientations, comparing pristine and irradiated (0.152~dpa) samples.}
    \label{fig: crack_length}
\end{figure}

Crack length evolution across all crystallographic orientations is presented in Fig.~\ref{fig: crack_length}. The crack length was estimated by projecting the evolving crack surface onto the propagation ($x$-$z$) plane and computing the mean projected crack length along the propagation direction, which minimizes sensitivity to local crack-front deviations across the specimen thickness. In the pristine samples, crack growth is most limited for the (011) orientation because its geometry favors activation of two intersecting \{111\} slip systems, producing localized plasticity, stacking faults, and early twinning at the crack tip. This concentrated deformation blunts the crack and efficiently shields the crack tip. In contrast, the (001) orientation exhibits limited slip activity, and the (111) orientation distributes plasticity across multiple planes rather than concentrating it near the crack tip, resulting in longer cracks. The strongest irradiation effect is observed for the (011) orientation, where restricted slip activity, combined with void absorption at the crack front, promotes accelerated crack advance. The (001) samples show nearly identical crack extension in pristine and irradiated states, but with a relatively fast crack growth due to limited plasticity. For the (111) orientation, crack propagation remains strongly suppressed, irrespective of irradiation, as the efficient activation of multiple slip systems sustains plastic redistribution and blunts the crack tip.
\subsection{Traction--separation analysis}
\vspace{0.2cm}
To further quantify the effect of irradiation on fracture resistance and to complement the global stress--strain analysis, a T--S law is used to characterise the fracture behavior at the atomic scale. The T--S curve, widely used to analyse fracture processes in MD simulations \citep{Krull2011, barrows2016traction, ding2020multi}, defines the relationship between the normal traction transmitted across the crack surfaces and the corresponding crack opening displacement. In this framework, fracture is described as a progressive reduction of cohesive traction with increasing crack opening displacement.

Due to the finite size of the simulation cell, extended defects such as stacking faults may interact with system boundaries and, in some cases, re-enter the simulation domain, leading to an apparent increase in their global density. This is a general feature of finite atomistic models and may occur for all analyzed orientations, as can be seen in Fig.~\ref{fig:purecompare}. To mitigate these artifacts, the T--S analysis is performed on local subvolumes located within regions of stable crack propagation and sufficiently distant from the boundaries, ensuring that the extracted T--S response reliably reflects the local crack-tip deformation mechanisms.

In our simulations, five distinct regions were distinguished near the crack tip, as shown in Section~\ref{sec:Section2}, Fig.~\ref{fig: Mode-i}. A consistent and physically meaningful evaluation of the T--S response requires selecting representative subvolumes along the crack path \citep{barrows2016traction, ding2020multi, ZHU2021102999}, from which local traction and crack opening displacement are obtained. The normal traction component is obtained by averaging the stress distribution within two adjacent subvolumes: the upper region (n), and the lower region (n\textit{$^\prime$}), positioned on opposite sides of the crack plane (compare Fig. \ref{fig: Mode-i}). The crack opening displacement is determined by computing the relative displacement between these two regions, specifically by subtracting the average atomic displacement of the upper region (n) from that of the lower one  (n\textit{$^\prime$}).
\begin{figure}[t!]
    \centering
    \includegraphics[scale=0.35]{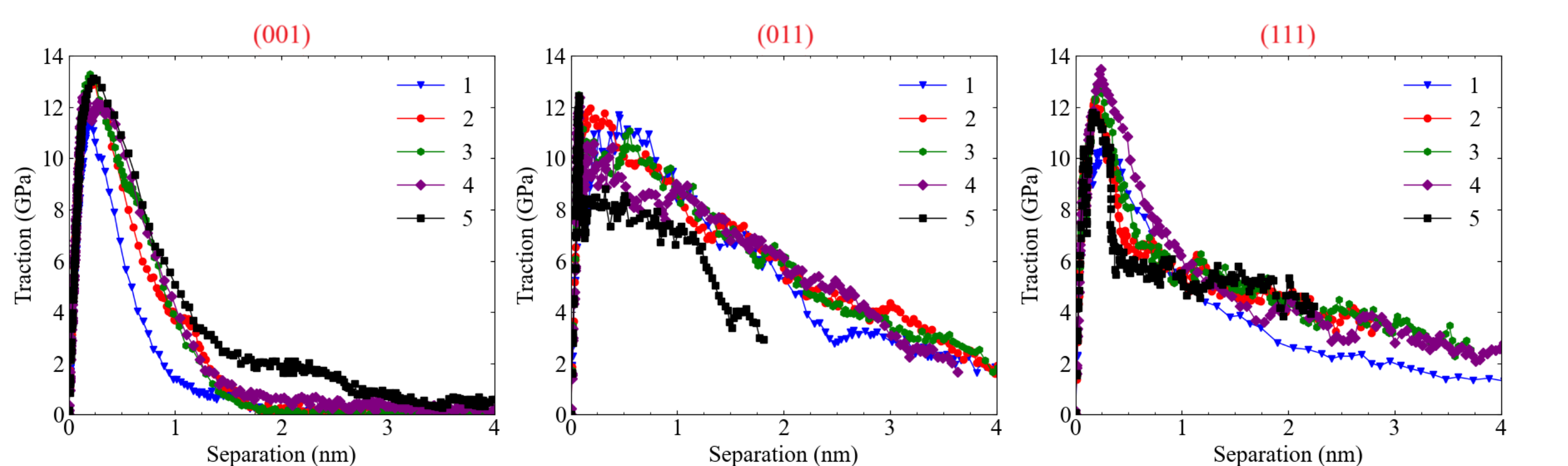}
    \caption{T--S curves for pristine samples in the (001), (011), and (111) orientations, derived from discrete sub-volumes (boxes 1-5) along the crack propagation path.}
    \label{fig: filter}
\end{figure}
\begin{figure}[b!]
    \centering
    \includegraphics[scale=0.275]{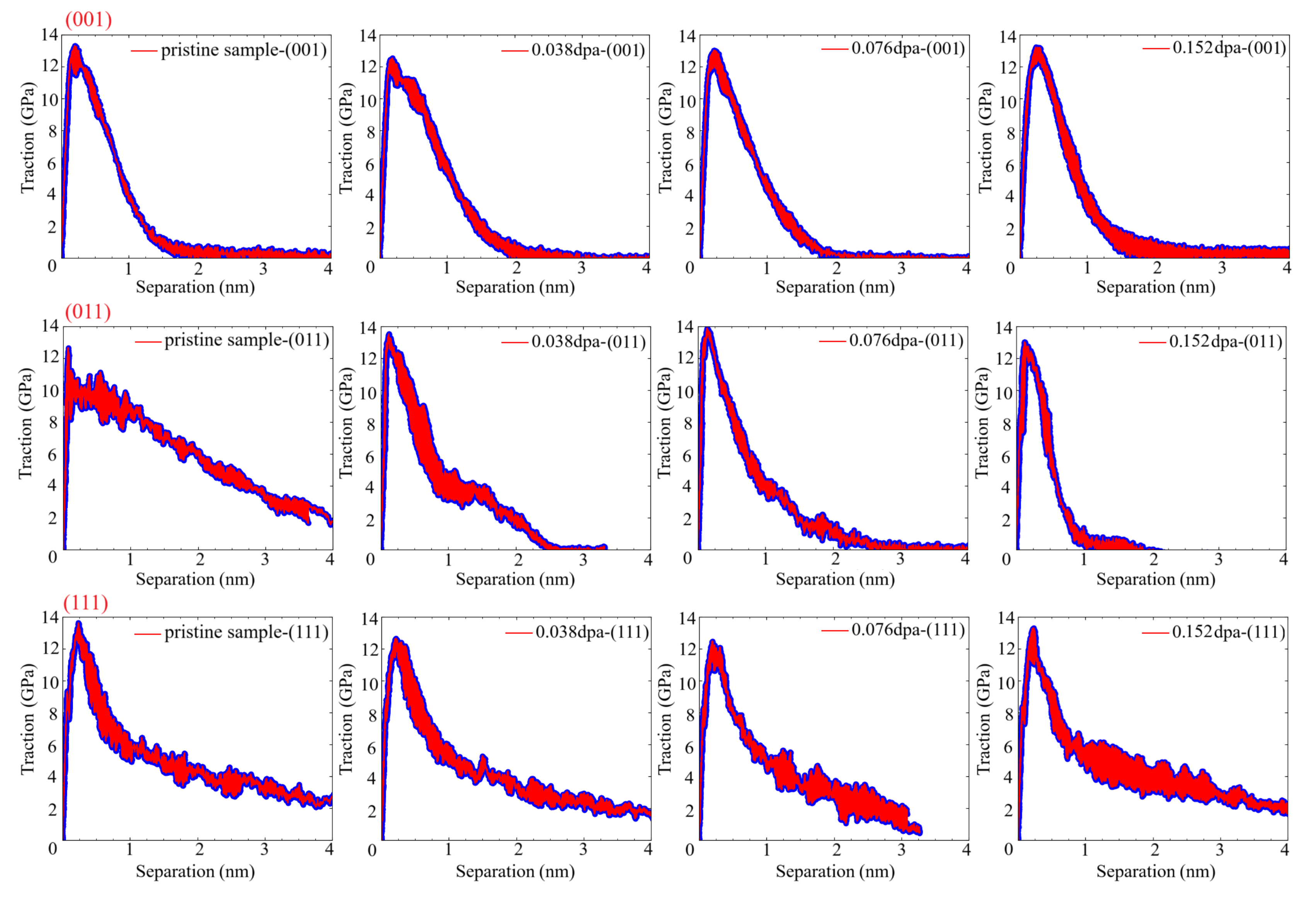}
    \caption{T--S responses extracted from two representative subvolumes (boxes 3 and 4) identified as representative regions along the crack path. Columns correspond to increasing irradiation levels, from pristine to 0.152~dpa; rows show orientation-dependent behavior: (001) (top), (011) (middle), and (111) (bottom).}
    \label{fig: TSall}
\end{figure}

Among the five subvolumes defined along the crack path (see Fig.~\ref{fig: Mode-i}), boxes 3 and 4 are selected for further analysis based on the T--S curves shown in Fig.~\ref{fig: filter}, as they correspond to regions of relatively stable crack propagation and exhibit well-defined separation behavior, including a clear peak and post-peak softening.
Boxes 1 and 2 are excluded because they are located too close to the crack tip, where rapid crack advance results in artificially low traction values. Box 5 corresponds to a region behind the active crack front, where defect accumulation and dislocation activity can distort the local mechanical response. For these reasons, boxes 3 and 4 provide the most reliable basis for quantifying orientation-dependent fracture behavior.

Fig.~\ref{fig: TSall} presents the T--S responses for all analyzed crystallographic orientations, extracted from boxes 3 and 4. Each row corresponds to a specific crystallographic orientation, with curves for increasing irradiation levels arranged from left to right within the row.

The (011) orientation (second row in Fig.~\ref{fig: TSall}) exhibits a pronounced DBT under irradiation. With increasing dose, the pre-failure crack opening decreases and the post-peak T--S slope steepens, reflecting constrained plastic dissipation and enhanced crack driving due to radiation-induced defects. The (001) orientation (first row in Fig.~\ref{fig: TSall}) shows minimal changes in maximum traction and failure displacement. The (111) orientation (third row in Fig.~\ref{fig: TSall}) maintains a relatively stable T--S response, with gradual post-peak softening. Interestingly, the maximum traction ($T_{\rm max}$), a key T--S parameter, does not consistently correlate with irradiation. Consequently, post-peak softening and total separation displacement offer a more reliable metric for assessing radiation-induced embrittlement \citep{XIE2023}.

\subsection{Energy-based assessment of radiation-induced embrittlement}
\label{sec: Section4.3}
\vspace{0.2cm}
In this subsection, the atomic-scale fracture energy (\( g_{\rm f} \)) is evaluated as a quantitative measure of fracture resistance, obtained directly from the T--S response of small subvolumes at the crack tip. 

This energetic approach is related to the extended Griffith model for fracture in ductile materials \citep{orowan1945notch, JOKL1980, BELTZ1992}, in which the critical energy release rate (\( G_{\rm c} \)) is expressed as a sum of the energy associated with surface formation, $\gamma_{\rm s}$, and the the plastic work per unit crack area resulting from dislocation activity, $w_{\rm p}$, 
such that $G_{\rm c} = 2\gamma_{\rm s} + w_{\rm p}$. Due to the finite size of the simulated volumes, MD captures only a small fraction of the crack-tip plastic zone, so that the plastic work contribution is underestimated and the resulting $G_{\rm c}$ values cannot be interpreted as representative macroscopic quantities. 
However, the MD-based estimates of $G_{\rm c}$ may serve as qualitative indicators of some physical phenomena, e.g., the DBT induced by temperature \citep{LIN2025} or by irradiation embrittlement \citep{Ustrzycka2025}.

\begin{figure}[t!]
    \centering
    \includegraphics[scale=0.43]{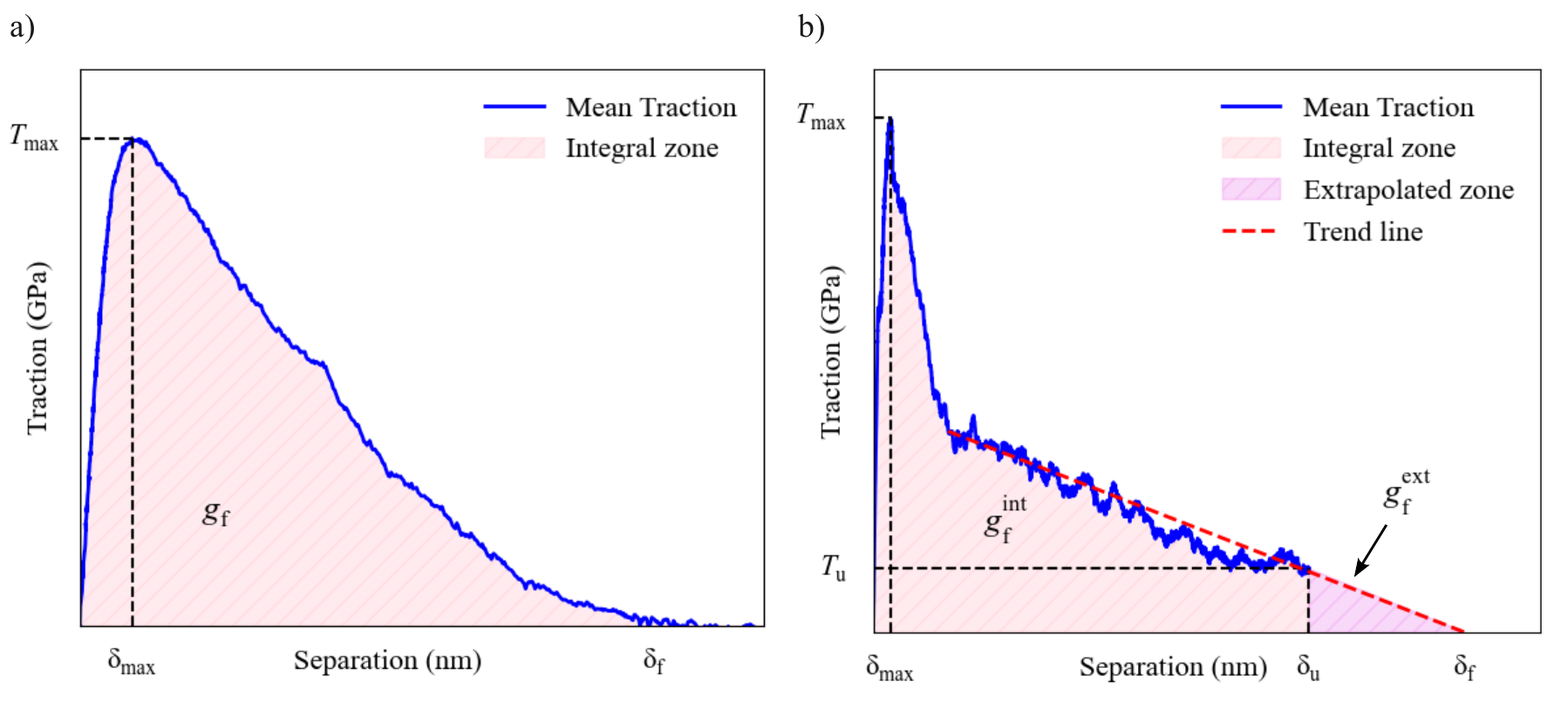} 
\caption{Schematic representation of the fracture energy (\( g_{\rm f} \)) obtained from traction--separation behavior. (a)~Complete T--S curve; the blue line shows the mean traction response, and the shaded pink area corresponds to the integrated fracture energy (\( g_{\rm f} \!=\! g_{\mathrm{\rm f}}^{\mathrm{int}}\)). (b) Incomplete T--S curve requiring extrapolation; the solid blue line represents the computed mean traction, the light pink area indicates the integrated portion (\( g_{\mathrm{\rm f}}^{\mathrm{int}} \)), and the red dashed line shows the linear extrapolation used to estimate the remaining contribution (\( g_{\mathrm{\rm f}}^{\mathrm{ext}} \), dark pink area), so that \( g_{\rm f} \!=\! g_{\mathrm{\rm f}}^{\mathrm{int}} \!+\! g_{\mathrm{\rm f}}^{\mathrm{ext}} \).}
 \label{fig: Gcr-method}
\end{figure}

In our work, \sts{the atomic-scale fracture energy} $g_{\rm f}$ is defined as the area under the T--S curve, see Fig.~\ref{fig: Gcr-method}, and represents the local interface formation energy required to open a crack in the MD subvolumes. Unlike the continuum-scale $G_{\rm c}$, which accounts for both surface formation and bulk plastic dissipation, $g_{\rm f}$ predominantly reflects the energy of creating new fracture surfaces, with only a minor contribution from plastic work due to the limited portion of the crack-tip plastic zone covered by the MD subvolumes.

The T--S responses shown in Fig.~\ref{fig: TSall} for the (111) orientation and the unirradiated the (011) sample retain nonzero tractions at the maximum separation. This indicates that in these cases, the crack propagation was arrested before reaching complete material separation. This behavior is attributed to progressive crack tip blunting, where intensive dislocation activity near the crack front inhibits further crack opening. As a result, the separation process remains incomplete. To ensure a consistent, energy-based assessment across all configurations, an extrapolated contribution to the energy release rate, denoted $g_{\mathrm{\rm f}}^{\mathrm{\rm ext}}$, is introduced for cases in which T--S response does not reach zero traction within the simulated separation range. The terminal part of the T--S curve is extended by a linear extrapolation from the last available data point to an assumed final separation, $\delta_{\rm f}$, at which the traction vanishes. The simple linear extrapolation adopted here is sufficient for the purpose of illustrating the effect of radiation defects on fracture energy $g_{\rm f}$. Alternative nonlinear extrapolations introduce only small quantitative changes that do not alter the overall picture. This procedure, illustrated in Fig.~\ref{fig: Gcr-method}(b), enables estimation of the energy associated with the unobserved segment of the response, corresponding to the final phase of material separation not captured in the simulation. The fracture energy, (\( g_{\rm f} \)), is then composed of two contributions: the numerically integrated part of the curve, $g_{\mathrm{\rm f}}^{\mathrm{\rm int}}$, and the extrapolated segment, $g_{\mathrm{\rm f}}^{\mathrm{\rm ext}}$, according to equation: $g_{\rm f} = g_{\mathrm{\rm f}}^{\mathrm{\rm int}} + g_{\mathrm{\rm f}}^{\mathrm{\rm ext}}$.

\begin{figure}[t!]
    \centering
    \includegraphics[scale=0.66]{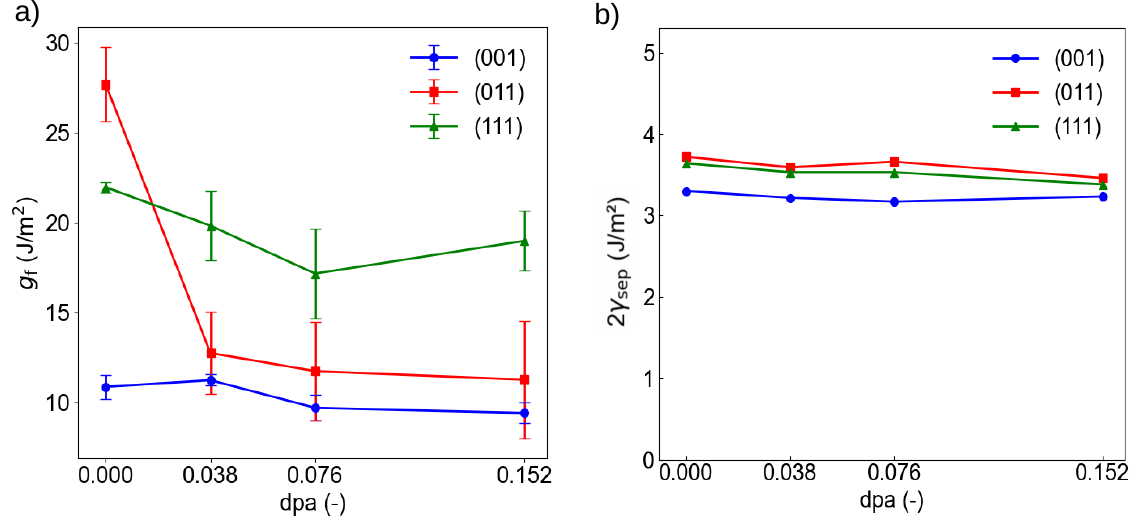}
    \caption{(a) Fracture energy, \( g_{\rm f} \), and (b) separation energy, 2$\gamma_{\rm sep}$, as a function of irradiation dose for different orientations.}
    \label{fig: Gcr}
\end{figure}

Fig.~\ref{fig: Gcr} shows in panel (a) the fracture energy, $g_{\rm f}$, and in panel (b) the separation energy, $2\gamma_{\rm sep}$, calculated for the three orientations as a function of the irradiation dose.

The separation energy, $2\gamma_{\rm sep}$, represents the energy required to break atomic bonds and create two planar free surfaces in the absence of plastic deformation. It thus provides a lower bound for the energy of surface formation in a fracture process, $2\gamma_{\rm s}$, which is the dominant contribution to $g_{\rm f}$. The nearly constant values of $2\gamma_{\rm sep}$ across irradiation levels confirm that radiation defects have little effect on the intrinsic surface energy. Moreover, for both the (011) and (111) orientations, the free surface (nominal fracture surface) is a \{011\}-type plane, see Table \ref{Tab:orientation}, so the respective separation energies should be identical, which is confirmed in Fig.~\ref{fig: Gcr}(b).

Since $2\gamma_{\rm sep}$ remains essentially unchanged, the radiation-induced reduction of $g_{\rm f}$ reflects the progressive suppression of plastic dissipation, $w_{\rm p}$, in the MD subvolumes at the crack tip. Although $w_{\rm p}$ is only a minor contribution due to the limited extent of the captured plastic zone, its reduction with increasing dpa provides a meaningful indicator of radiation-induced embrittlement. 

Among the analyzed orientations (see Fig.~\ref{fig: Gcr}(a)), the (011) orientation exhibits the most pronounced decrease in $g_{\rm f}$, pointing to a strong suppression of crack-tip plasticity. Before irradiation, coordinated dislocation emission, stacking-fault formation, and twinning generate crack-tip blunting and plastic dissipation. After irradiation, radiation-induced defects (SFTs, loops, and voids) act as pinning centers, blocking stacking-fault propagation, limiting twin formation, and promoting immobile junctions and void coalescence, which localize deformation and reduce the plastic contribution to $g_{\rm f}$.

The (001) orientation consistently shows the lowest fracture energies, while the (111) orientation maintains the highest resistance to fracture under irradiation.
Whereas the relatively small changes in $g_{\rm f}$ between 0.038 and 0.152~dpa indicate that crack-tip plasticity suppression in the MD subvolumes reaches a steady state, this does not imply saturation of macroscopic embrittlement. 
Note that the actual values of $g_{\rm f}$ are not central to conclusions, the radiation-induced \emph{changes} in $g_{\rm f}$ are the relevant quantity, and for this purpose the linear extension of the T--S curve, shown in Fig.~\ref{fig: Gcr-method}, serves the purpose well.

Importantly, the quantitative energy-based analysis in this work is not based on global measures of dislocation density or stacking-fault content, but on selected subvolumes along the crack path that correspond to regions of stable crack propagation. These regions are chosen to reduce the influence of system-size effects and defect reflection, ensuring that the extracted fracture energies reflect the local crack-tip deformation mechanisms.

For each irradiation level,  \( g_{\rm f} \) values were obtained from several statistically independent realizations. The reported results represent mean values with corresponding standard deviations, based on four realizations for the (011) orientation and three for (001) and (111).

%% file: sections/5_Summary.tex
\section{Conclusions}
\label{sec:Section6}

This study provides a comprehensive atomistic perspective on radiation-induced embrittlement in fcc $\rm Fe_{55}Ni_{19}Cr_{26}$ alloy, revealing how crystallographic orientation mediates fracture behavior through its influence on plasticity and defect interactions.
By combining MD simulations with orientation-resolved traction--separation analysis, we establish a mechanistic framework that links lattice geometry, radiation defect structures, and fracture energy dissipation.
The study shows that radiation-induced embrittlement in the alloy is governed by the interplay between crystallographic orientation and defect evolution. Through MD simulations combined with traction--separation analysis, we reveal that fracture resistance results from a balance between defect accumulation, radiation-induced alterations to deformation processes, and the orientation-dependent capacity of the lattice to accommodate plasticity and crack propagation. 

Fracture simulations reveal that the ductile-to-brittle transition manifests through distinct mechanisms in different orientations. In the (001) orientation, limited slip activity constrains plastic relaxation regardless of irradiation level, leading to consistently low fracture energy. In contrast, the (011) orientation exhibits a sharp reduction in $g_{\rm f}$, driven by defect-induced obstruction of stacking fault propagation and formation of immobile dislocation junctions, such as Lomer--Cottrell locks. The (111) orientation, by activating multiple slip systems and efficiently incorporating defects into stacking faults, sustains high levels of energy absorption even in irradiated conditions, resisting the transition to brittle fracture. All the above conclusions pertain to the relatively low density of radiation-induced defects that can be achieved in our MD setting.

The orientation-dependent variation of the fracture energy ($g_{\rm f}$) obtained from the T--S analysis underscores that fracture resistance under irradiation is controlled by the interplay between crack front and slip geometry and the spatial distribution of defects, which together govern energy dissipation and crack advance.

Taken together, these results emphasize that radiation-induced fracture in alloys is governed by orientation-sensitive mechanisms of plastic deformation and defect interaction. The directional nature of slip activity, the crystallographic dependence of defect accommodation, and the evolution of separation behavior collectively determine the material's fracture resistance under irradiation. By linking atomic-scale mechanisms with continuum-scale fracture metrics, this work provides a physically grounded perspective on embrittlement.

The principal advancement of this study lies in establishing a unified, atomistically based framework that connects radiation-induced defect structures with the orientation-dependent evolution of crack-tip plasticity and fracture energy. By identifying how defect–mechanism interactions modulate plastic dissipation across different lattice orientations, the model offers a mechanistic basis for interpreting irradiation-driven embrittlement beyond phenomenological descriptions. This framework provides quantitatively grounded criteria for distinguishing orientation-sensitive DBT pathways and clarifies the fundamental role of slip and twinning geometry in governing fracture resistance under irradiation.

The results contribute to a mechanistic understanding of irradiation-induced ductile-to-brittle transitions in austenitic steels and provide physically grounded guidance for multiscale modeling under service-relevant irradiation conditions.